\documentclass[reprint,amsmath,amssymb,aip,
journal=rmp]{revtex4-2}
\usepackage{dcolumn} 
\usepackage{graphicx} 
\usepackage[version=3]{mhchem}
\usepackage{color,soul}
\usepackage{multirow}
\usepackage{braket}
\usepackage{xcolor}
\usepackage{bm}
\usepackage{comment}
\usepackage{hyperref}
\usepackage{physics}   
\usepackage[normalem]{ulem}

\newcommand{\hbm}[1]{\hat{\bm{#1}}} 
\newcommand{\pp}[2]{\frac{\partial{#1}}{\partial{#2}}}

\newcommand{\hbGamma}{{\hat{\bm \Gamma}}}
\newcommand{\blambda}{{\bm \lambda}}
\newcommand{\hH}{\hat H}
\newcommand{\hT}{\hat T}
\newcommand{\hO}{\hat O}
\newcommand{\hP}{\hat P}

\newcommand{\hU}{\hat U}
\newcommand{\hV}{\hat V}

\newcommand{\hbp}{{\hat{\bm p}}}
\newcommand{\hbl}{{\hat{\bm l}}}
\newcommand{\hbs}{{\hat{\bm s}}}
\newcommand{\hbX}{{\hat{\bm X}}}
\newcommand{\hbP}{{\hat{\bm P}}}

\newcommand{\hbx}{{\hat{\bm x}}}

\newcommand{\bA}{{\bm A}}
\newcommand{\bP}{{\bm P}}
\newcommand{\bDelta}{{\bm \Delta}}
\newcommand{\bPi}{{\bm \Pi}}
\newcommand{\bmm}{{\bm m}}
\newcommand{\bmu}{{\bm \mu}}
\newcommand{\bX}{{\bm X}}
\newcommand{\bx}{{\bm x}}

\newcommand{\bI}{{\bm I}}
\newcommand{\bV}{{\bm V}}

\newcommand{\bB}{{\bm B}}

\newcommand{\bp}{{\bm p}}

\newcommand{\bd}{{\bm d}}

\begin{document}

\title{The Phase-Space Way To Electronic Structure Theory and Subsequently Chemical Dynamics}

\author{Xuezhi Bian}
\email{xzbian@princeton.edu}
\affiliation{Department of Chemistry, Princeton University, Princeton, New Jersey 08544, USA}
\author{Titouan Duston}
\email{td1272@princeton.edu}
\affiliation{Department of Chemistry, Princeton University, Princeton, New Jersey 08544, USA}
\author{Nadine Bradbury}
\email{nadinebradbury@princeton.edu}
\affiliation{Department of Chemistry, Princeton University, Princeton, New Jersey 08544, USA}
\author{Zhen Tao}
\email{ct1185@princeton.edu }
\affiliation{Department of Chemistry, Princeton University, Princeton, New Jersey 08544, USA}
\author{Mansi Bhati}
\affiliation{Department of Chemistry, Princeton University, Princeton, New Jersey 08544, USA}
\author{Tian Qiu}
\affiliation{Department of Chemistry, Princeton University, Princeton, New Jersey 08544, USA}
\author{Xinchun Wu}
\affiliation{Department of Chemistry, Princeton University, Princeton, New Jersey 08544, USA}
\author{Yanze Wu}
\affiliation{Department of Chemistry, Northwestern University, Evanston, Illinois, 60208, USA}
\author{Joseph E. Subotnik}
\email{subotnik@princeton.edu}
\affiliation{Department of Chemistry, Princeton University, Princeton, New Jersey 08544, USA}

\date{\today}

\begin{abstract}
    Phase-space electronic structure theory offers up a new and powerful approach for tackling problems with coupled nuclear-electronic dynamics in a fashion that goes beyond Born-Oppenheimer (BO) theory. Whereas BO theory stipulates that we consider electronic states parameterized by nuclear position  $\bX$ only, i.e. molecular orbitals are functions of nuclear positions but not nuclear velocities, phase-space (PS) theory allows for electronic states to be parameterized by both nuclear position $\bX$ and nuclear momentum $\bP.$ As a result, within a phase-space approach, one can directly recover many new features, including  electronic momentum and vibrational circular dichroism spectra. Moreover, phase-space electronic structure theory is exact for the hydrogen atom and, for a set of model problems, the method can even improve upon vibrational energies relative to BO theory. Perhaps most importantly, the phase-space approach offers up a very new perspective on spin physics, stipulating that molecules and materials with degenerate or nearly degenerate ground states (especially due to spin degeneracy) display broken-symmetry ground states in their phase-space potential energy surfaces. This last feature opens up novel possibilities for exploring spin chemistry (including the Einstein-de Haas effect and chiral induced spin selectivity) using  well established electronic structure theory methods.  At the end of the day,
   in order to tackle a host of exciting electronic dynamical phenomena, especially subtle problems in magnetic chemistry, it will be essential for the electronic structure community to pivot towards diagonalizing $\hH_{\rm PS}(\bX,\bP)$ rather than $\hH_{\rm BO}(\bX)$. 
   
\end{abstract}

\maketitle

\noindent {\bf Notation:}  In what follows, nuclei  are indexed by capital $I,J$ and (for index $I$) have coordinates   $(\bX_I,\bP_I)$ charge $Q_I$ and mass $M_I$. Electrons and electronic states  are  indexed by lower case  $i,j,k,\ldots$ and correspond to coordinates $(\bx_i,\bp_i)$, with charge $-e$ and mass $m_e$. Hats indicate operators (e.g. $\hH$) and boldface (e.g. $\bX$) indicates three-dimensional vectors or tensors.

\section{Born-Oppenheimer Theory}
\label{sec-BO}
The Born-Oppenheimer framework is the cornerstone of chemistry\cite{born1985quantentheorie, born1996dynamical}.   Based on the mass difference between nuclei and electrons, the framework stipulates that the electronic states of any molecular or material system can be parameterized by the positions $\bX$ (and not momentum $\bP$) of the nuclei\cite{cederbaum:review:conicalbook}. In the simplest terms, a modern electronic structure package takes as input the nuclear coordinates and  in turn spits out the electronic orbitals; a standard Q-Chem hartree-fock calculation\cite{qchem6} never requests as input the momenta (or even the masses) of the nuclei.

For the moment, let us ignore  fine-structure and spin effects as well as external fields (see Sec. \ref{sec-spin} for a discussion of spin and Sec. \ref{sec-mag} for a few remarks about magnetic fields). 
Mathematically, for a system of $N_e$ electrons and $N_n$ nuclei,  the total Hamiltonian is of the form (setting $e>0$):
\begin{eqnarray}
\label{eq:Htot}
    \hH_{tot} &=&     \sum_{I=1}^{N_n} \frac{\hbP_I^2}{2M_I} + \sum_{i=1}^{N_e} \frac{\hbp_i^2}{2m_e} 
    -    \sum_{I=1}^{N_n} \sum_{i = 1}^{N_e} \frac{Q_Ie} {4\pi \epsilon_0 \left| \hbx_i - \hbX_I \right|} \nonumber \\
    & & \; 
    +    \sum_{I=1}^{N_n} \sum_{J = I+1}^{N_n} \frac{Q_I Q_J}{4\pi \epsilon_0 \left| \hbX_J - \hbX_I \right|}  \nonumber \\
     & & \;
    + \sum_{i=1}^{N_e} \sum_{j = i+1}^{N_e} \frac{e^2}{4\pi \epsilon_0 \left| \hbx_i - \hbx_j \right|}. 
\end{eqnarray}

According to the BO framework\cite{yarkony:review:conicalbook}, the natural way to model such a system is to decompose the total Hamiltonian into a nuclear kinetic energy ($\hat{T}_{\rm n}$) term and an electronic Hamiltonian ($\hH_{\rm el}$):
\begin{eqnarray}
\label{eq:Hel}
    \hH_{\rm tot} & =& \hT_{\rm n} + \hH_{\rm el} \\
     \hT_{\rm n} &=& \sum_{I=1}^{N_n} \frac{\hbP_I^2}{2M_I} \\
     \hH_{\rm el} &=& \sum_{i=1}^{N_e} \frac{\hbp_i^2}{2m_e} 
    -    \sum_{I=1}^{N_n} \sum_{i = 1}^{N_e} \frac{Q_Ie} {4\pi \epsilon_0 \left| \hbx_i - \hbX_I \right|} \nonumber \\
    & & \;  
    +    \sum_{I=1}^{N_n} \sum_{J = I+1}^{N_n} \frac{Q_I Q_J}{4\pi \epsilon_0 \left| \hbX_J - \hbX_I \right|} \nonumber \\ 
     & & \;   
    + \sum_{i=1}^{N_e} \sum_{j = i+1}^{N_e} \frac{e^2}{4\pi \epsilon_0 \left| \hbx_i - \hbx_j \right|}. 
\end{eqnarray}
Thereafter, one diagonalizes the electronic Hamiltonian and generates eigenvalues $V_k^{\rm BO}(\bX)$  and eigenvectors $\Phi_k(\bx;\bX)$ that depend parametrically only on $\bX$ (and not $\bP$):
\begin{eqnarray}
\label{eq:Hel_diag}
    \hH_{\rm el}(\bX) \ket{\Phi_k} = V_k^{\rm BO}(\bX) \ket{\Phi_k}. 
\end{eqnarray}
Finally, if one considers the total Hamiltonian in the basis of adiabatic states, the result is:
\begin{eqnarray}  \label{eq:Hadfull}
     \hH_{\rm tot}' &=&  
    \sum_{mjkI} \ket{\Phi_j} \frac {({\bm \hP}_I \delta_{jm}  - i\hbar{\bm d}^I_{jm}) \cdot ({\bm \hP}_I\delta_{mk} - i\hbar{\bm d}^I_{mk})} {2M_I} 
    \bra{\Phi_k} \nonumber \\
    & & \; 
    + \sum_j V_j^{\rm BO}(\hbX) \ket{\Phi_j}\bra{\Phi_j},
\end{eqnarray}
where $\bd_{jk}$ is the matrix of derivative couplings,
\begin{eqnarray}
    \bd^I_{jk} = \bra{\Phi_j}  \frac{\partial}{\partial \bX_I} \ket{\Phi_k},
\end{eqnarray}
to be discussed below.  Eq. \ref{eq:Hadfull} can be derived in many ways, most easily by using an explicit unitary transformation $\hU(\bX)$
from the position basis to the basis of electronic states. Mathematically, suppose we diagonalize $\hH_{el}$, i.e. set
$\hH_{\rm el} =  \hU \hV^{\rm BO} \hU^{\dagger}$, so that the $\hU$ transformation defines the BO framework.
If we then conjugate the total Hamiltonian in Eq. \ref{eq:Htot} by $\hU^{\dagger}$, the result is 
\begin{eqnarray}
\label{eq:BO1}
\hH_{\rm tot}' &=& \hU^{\dagger} \hH_{\rm tot} \hU  \\
\label{eq:BO2}
 &=&     \sum_{I} 
 \frac{1}{2M_I} \left( {\bm \hP}_I - i\hbar\hU^{\dagger} \frac{\partial \hU }{\partial \bX_I} \right)^2 + 
 \hV^{\rm BO}.
\end{eqnarray}
Eq. \ref{eq:BO2} is equivalent to Eq. \ref{eq:Hadfull} if we identify $\left(\hU^{\dagger} {\partial \hU }/{\partial \bX_I} \right)_{jk}$ as the derivative coupling $\bd_{jk}^I$.

Up until now, everything has been exact and we have merely followed the BO {\em framework}. Hereafter, the BO {\em approximation} arises by ignoring the derivative  couplings in Eq. \ref{eq:Hadfull}. For instance, with classical nuclei, the relevant BO equations of motion along surface $k$ are simply:
\begin{eqnarray}
\label{eq:xdot:bo}
    \dot{\bX}_I &=& \frac{\partial \langle\hat{H}'\rangle}{\partial \bP_I} = \frac{\bP_I}{M_I}, \\
    \label{eq:pdot:bo}
    \dot{\bP}_I &=&  -\frac{\partial \langle\hat{H}'\rangle}{\partial \bX_I} = -\frac{\partial V_k}{\partial \bX_I}.  
\end{eqnarray}
Before analyzing the implications of this BO approximation (see Secs. \ref{sec-BO-fail1} and \ref{sec-BO-fail2} below), a few words are appropriate regarding the derivative coupling (a matrix of vectors)
that incorporates all non-BO effects. In particular, note that $\bd$ satisfies the following four equations\cite{yanzewu:2024:jcp:pssh_conserve}:
\begin{eqnarray}
    -i\hbar\sum_{I} {\bm d}^I_{jk} + \bra{\Phi_j}\bm{\hat{p}} \ket{\Phi_k} &=& 0,\label{eq:dconstrain1a}  \\
    \sum_I \bm \nabla_{I} {\bm d}^J_{jk}  &=& 0, 
    \label{eq:dconstrain2a}\\
    -i\hbar\sum_{I}{\bm X}_{I} \times {\bm d}^I_{jk} + \bra{\Phi_j} \bm{\hat{l}} + \bm{\hat{s}} \ket{\Phi_k} &=& 0,\label{eq:dconstrain3a}\\
     -\sum_I \left({\bm X}_I \times \bm \nabla_{I} {d}^{J\beta}_{jk}\right)_{\alpha} &
     =&  \sum_{\gamma} \epsilon_{\alpha \beta \gamma} d^{J \gamma}_{jk},\label{eq:dconstrain4a} 
\end{eqnarray}
where $\hat{\bm l}$ and $\hat{\bm s}$ are the electronic orbital and spin angular momentum operators.
In words, Eqs. \ref{eq:dconstrain1a} and \ref{eq:dconstrain3a} reflect the fact that the system is unchanged under translation and rotation of the entire system (of electrons and nuclei) and that the adiabatic electronic states are taken relative to the nuclear positions.  Eqs. \ref{eq:dconstrain2a} and \ref{eq:dconstrain4a} reflect that the derivative coupling vectors themselves transform correctly under translation and rotation. In other words, a total translation of the system  (by amount $\bDelta$) does not change the form of the derivative coupling; a total rotation of the system (by a rotational transformation $\mathcal{R}$) simply induces a rotation of the derivative couplings:
\begin{eqnarray}
\bd^I_{jk}(\bX + \bDelta) &=&  \bd^I_{jk}(\bX ) \\
\bd^I_{jk}( \mathcal{R}\bX) &=&  \mathcal{R}\bd^I_{jk}(\bX ) 
\end{eqnarray}
Another way to understand Eqs. \ref{eq:dconstrain1a}  and \ref{eq:dconstrain3a} is to note that, if one chooses the adiabatic electronic states relative to the nuclear coordinates, the following rules must hold:

\begin{widetext}
\begin{eqnarray}
\label{eq:final:1a}
    \left( \sum_j \frac{\partial}{\partial \bx_j} + 
        \sum_I \frac{\partial} {\partial \bX_I} \right) \Phi_k(\bx;\bX) &=& 0, \\
        \label{eq:final:3a}
    \left( \frac{i}{\hbar}\sum_i {\bm{\hat s}}_i  + \sum_i \bx_i \times \frac{\partial}{\partial \bx_i} + 
         \sum_I  \bX_I \times \frac{\partial} {\partial \bX_I} \right) \Phi_k(\bx;\bX) &=& 0 .
\end{eqnarray}
    
\end{widetext}
Eqs. \ref{eq:dconstrain1a}  and \ref{eq:dconstrain3a} follow by multiplying Eqs. \ref{eq:final:1a}  and \ref{eq:final:3a} on the left by $\Phi_j^*(\bx;\bX)$ and integrating over $\bx$.

\subsection{Benefits of the Approach}
Before we proceed (inevitably) to enumerate the problems of the BO approximation, it is worth noting that for many many chemical processes (and by far the vast majority of ground state effects), the BO approximation works incredibly well. Indeed, almost all modern (classical) molecular dynamics simulations are based on the BO approximation,\cite{car1985unified} and biomolecular dynamics \cite{amber:2005} has been very successful at predicting the behavior of proteins when performing catalysis (resulting in the 2013 Nobel prize)\cite{warshel1976theoretical}.  As another example, within  the realm of exact quantum mechanics, one computes vibrational energies within the BO approximation often with quantitative results\cite{bernathbook}. At present, for solving vibrational spectroscopy accurately, one would normally presume  the biggest problems are usually (i)  solving for electronic wavefunctions with enough electron-electron correlation (which requires high level electronic structure theory\cite{tew:gauss:2007:ccsdt_basis_vibrational_spectroscopy}); and (ii) solving the nuclear Schrodinger equation and going beyond a harmonic approximation\cite{estrin:vibrational_review_2020} (e.g. with quantum monte carlo\cite{suhm:quantum_monte_carlo_review_1991} or centroid\cite{jang:199:jcp_voth} or ring polymer molecular dynamics (RPMD)\cite{mano:2004:jcp_rpmd1}). In general, or at least for many many systems, going beyond the BO approximation is not considered the highest priority for extracting vibrational energies. 


\subsection{Well Appreciated Drawbacks to the BO Approximation: ``Static Correlation''}
\label{sec-BO-fail1}
\begin{figure}
    \centering 
\includegraphics[width=\linewidth]{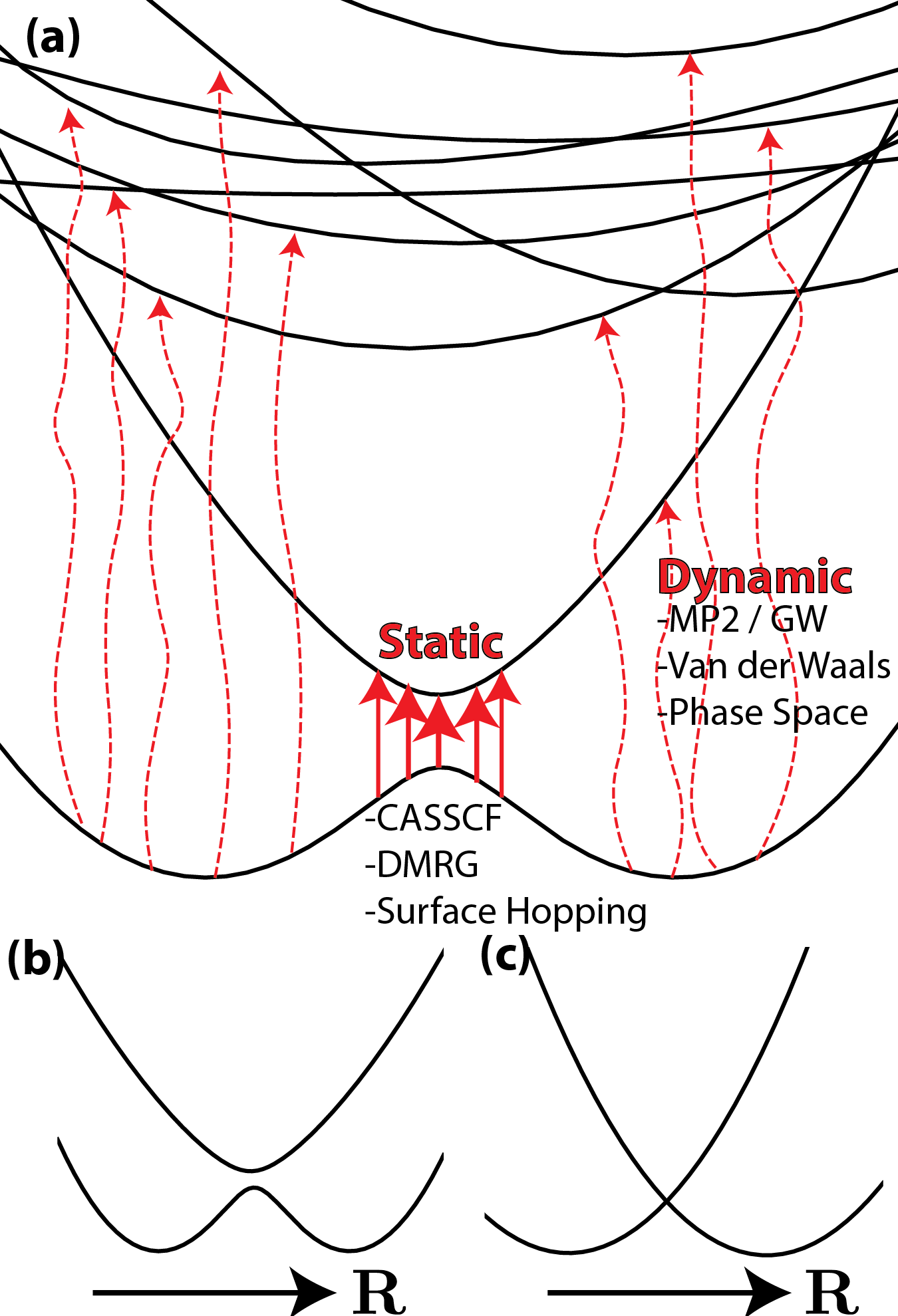}
    \caption{(a) A schematic view of   BO potential energy surfaces for a realistic molecule or material.  On the one hand, where the ground-state is energetically far removed from the excited states, one speaks of ``dynamical electron-nuclear correlation.''  At these geometries, corrections to BO theory require input from many excited states and lead to phase-space potential energy surfaces.  On the other hand, where the ground state is close to the excited state (i.e. at a curve crossing or conical intersection), one speaks of ``static electron-nuclear corelation''. In such cases, the nuclei do not know which surface to walk along, and one must propagate with Ehrenfest or surface-hopping dynamics.
    (b) Marcus theory assumes that the correct physics can be extracted from only two adiabatic states, and furthermore, (c) those two adiabatic states arise from two exactly diabatic states. }
    \label{fig_problems}
\end{figure}

There are well-known classes of problems where the BO approximation fails. Most famously, 
BO dynamics fail to describe dynamics through curve crossings and conical intersections\cite{nitzanbook}, e.g. all photochemical pathways \cite{nelson2020non}, as depicted heuristically in Fig. \ref{fig_problems}.  To understand this failure, note that using Hellman-Feynman theory and  the orthogonality of two eigenvectors, it follows that (if $V_j \ne V_k$)\cite{yarkony:review:conicalbook}:
\begin{eqnarray}
\label{eq:derv_coupl_hellman_feynman}
\bd_{jk}^I = \frac{\bra{\Phi_j} \partial \hH_{\rm el}/\partial \bX_I   \ket{\Phi_k}} {V_k - V_j}.
\end{eqnarray}
The end result of this analysis is that BO theory fails when two adiabatic surfaces come close together (through a curve crossing) and/or touch (at a conical intersection seam). In such a case, dynamics do not simply flow along one potential energy surface but rather wave packets split between the upper and lower surfaces.

Of course, the scenario above is very well understood within chemical dynamics and forms the basis for all theories of electronic relaxation, most famously the so-called ``Marcus theory of electron transfer''\cite{marcus:1956,marcus:1963,jortner:1999:advchemphys}). Consider Fig. \ref{fig_problems}(a). Here, we presume that dynamics in the bottom left well correspond to dynamics with an electron on a donor fragment; dynamics in the bottom right well correspond to dynamics with an electron on the acceptor fragment. 
Marcus theory assumes that, in this case, the ground state potential energy surface can be constructed from the lowest two adiabatic surfaces (Fig. \ref{fig_problems}(b)) and these two states are themselves rotations of two diabatic states (the Marcus parabolas [Fig. \ref{fig_problems}(c)]). For an {\em adiabatic} thermal electron transfer process (as depicted in Fig. \ref{fig_problems}(b), one imagines moving from the well on the left to the well on the right along the ground-state adiabat. 
Importantly, however, if the  gap between ground and excited state is small, then  the particle can leave the ground state surface when the dynamics pass through a crossing point (a {\em nonadiabatic} thermal electron transfer event). Mathematically, dynamics along the ground state become incorrect when the ground state changes character from donor to acceptor because the derivative coupling $\bd_{12}$  becomes large,
and from Eq. \ref{eq:derv_coupl_hellman_feynman}, this $\bd_{12}$  becomes large near a curve crossing when two state energies come close together.  In such regions, BO theory fails as a large off-diagonal coupling  drives a nuclear wavepacket from the ground state (surface 1) to the excited state (surface 2).

\subsubsection{Surface Hopping and Multireference Electronic Structure Theory}

All of the above has been well understood for many years in chemical physics\cite{barbara:1996:marcus}.
When curve crossings occur, the simplest model of electron transfer is to use transition state theory with an  electronic transmission factor, and there is today a robust and fairly complete understanding of electron transmission factors in the literature\cite{ripsjortner:1987, ripsjortner:1988, straub:1987:berne_nonadiabatic, jain:2015:friction}.
Moreover, theoretical chemistry has also spent several decades fervently going beyond transition state theory and developing practical dynamical algorithms to treat such ``well-known'' nonadiabatic problems that break BO theory. In particular, the picture above can be effectively quantified using 
 surface hopping approaches of various flavors \cite{landry:2012:marcus_afssh,reichman:2016:fssh,newton:chemrev}.
From a point of view inspired by John Tully\cite{tully:fssh,tully:faraday:fssh}, one often speaks of the system hopping between surfaces near such a curve crossing and/or intersection: the derivative coupling between two electronic states promotes radiationless electronic relaxation, which is a clear violation of BO dynamics.

Interestingly, it is worth mentioning that near the crossing point, not only do we find crucial hopping dynamics between two different surfaces, we also face a very difficult electronic structure problem.\cite{subotnik:2015:acr,lindh:2016:jpcl_benzophenone} At these geometries, the adiabatic electronic states are superpositions of two different  donor and acceptor diabatic electronic states, $\ket{\Phi} = \alpha \ket{D} + \beta \ket{A}$; thus the adiabatic states take on multireference character that cannot be captured easily with single determinant Hartree-Fock (HF) wavefunctions (or usually density functional theory [DFT])\cite{mhg:2005:chemrev} but rather require multiconfigurational SCF\cite{helgakerbook},   multireference configurational interaction(MRCI) \cite{werner:1988:mrci_internally_contracted,martinez:1996:jpc,lischka2018multireference} or other multireference methods\cite{olivares2015ab, booth2009fermion}.  In other words, 
at the crossing point, both the electron-electron problem and the electron-nuclear problem involve strong correlation between {\em two} different electronic states. Thus, henceforward, in analogy to the common practice in the electronic structure literature whereby the multireference character of a wavefunction at a crossing point reflects a problem of ``static electron-electron correlation'' (requiring a treatment beyond mean-field theory), we will describe the surface hopping problem at the curve crossing point as a manifestation of  ``static electron-nuclear correlation'' (require a treatment beyond BO theory). See the relevant ``static correlation'' arrow in Fig. \ref{fig_problems}(a). Both static correlation problems (electron-electron and electron-nuclear) are very well studied in chemistry.

\subsection{Slightly Less Appreciated Drawbacks to the BO Approximation: Dynamic Correlation}\label{sec-BO-fail2}
Whereas many chemists have long been interested in curve crossings, 
 most  chemists tend to ignore the failures of the BO approximation far from curve crossings, e.g. at the bottom of the wells  in Fig. \ref{fig_problems}.  In this region, indeed one need not worry about hopping  between surfaces. Nevertheless, 
 as well known to spectroscopists focusing on vibrational circular dichrosim (VCD) \cite{nafie:1983:jcp:el_momentum,nafie:1992:vcd,stephens:1985:jpcc_vcd,patchkovskii:2012:jcp:electronic_current},
 there are in fact important correlations missing from the BO approximation in this regime.  Here, the relevant correlations are {\em dynamic} and arise from the weak interaction of the state of  interest with many other high lying excited states. Note that this choice of nomenclature makes  direct analogy to those electronic structure methods (e.g. the Møller-Plesset perturbation [MP2] theory \cite{moller1934note}) that capture ``dynamic correlation''\cite{sherrillnotes, sulka:dyn_vs_stat_corr:2023:jctc, roca2012multiconfiguration}, i.e. the correlation that arises from the ground electronic state scattering weakly with higher electronic states  (e.g. through van der waals forces).
Note that the MP2 correction can yield a large post-hartree fock energy at the bottom of the well in Fig. \ref{fig_problems} (which is far from a curve crossing).

Far from a curve crossing, 
the most glaring failure of the BO approximation is the fact the BO approximation ignores electronic momentum (or, more precisely, it sets all electronic momentum to zero). This statement should be obvious to both beginner and expert readers. After all, according to BO theory, the electronic wavefunction for a hydrogen atom  in its ground state with nucleus at position $\bX$ is $\Phi_g(\bx;\bX) = C \exp(-(\bx - \bX)/a)$ where $a$ is the Bohr radius and $C$ is a constant.  What if a ground state hydrogen atom is moving with velocity $
\bV$? In such a case, the electronic wavefunction is unchanged according to BO theory; we simply replace $\bX$ by $\bX(t) = \bV t$. In both cases, if one evaluates the electronic momentum, the result is $\bra{\Phi_g} \hbp \ket{\Phi_g} = 0$, which follows because $\Phi_g(\bx)$ is real-valued; of course, the physical answer should be $m_e \bV$. The incorrect BO result extends to all wavefunctions, with many electrons and many nuclei.  According to BO theory, if a chemical system sits in its electronic ground state (or any stationary state for the matter),  the electronic angular and  linear momentum are  zero (even if the nuclei are moving).

Consider now the implications for computing VCD signals, where the rotational strength is given by:
\begin{eqnarray}
    \mathcal{R}_{12} = \mbox{Im} \left( \bmu_{12} \cdot {\bmm}_{21} \right).
\end{eqnarray}
Here, $\bmu_{12}$ and $\bmm_{21}$ are the electric and magnetic dipole moments (respectively)  between the initial (1) and final (2) states. 
In general, both $\bmu$ and $\bmm$ have electronic and nuclear parts\cite{nafie:1983:jcp:el_momentum, nafie:1997:arpc}, and both of these parts must be calculated precisely for a meaningful VCD spectra -- because the terms are usually of similar size but opposite sign (given the differences in charge between electrons and nuclei). However, note that, within BO theory, $\bmm_{12}$ is zero -- which precludes using BO theory directly to calculate VCD spectra.

Given this background, there have been a few approaches to fix up BO theory to allow for electronic momentum, including nuclear velocity perturbation (NVP)\cite{nafie:1992:vcd} and complete adiabatic (CA) perturbation\cite{nafie:1983:jcp:el_momentum} by Nafie.    Interestingly, suppose that one starts with an adiabatic electronic wavefunction $\ket{\Phi_j}$ and nuclear wavepacket $\ket{\chi_j}$, and one uses standard first order perturbation theory (with the  derivative coupling from Eq. \ref{eq:Hadfull} as the perturbation) to generate a corrected nuclear-electronic wavefunction $\ket{\Psi_j}$,
\begin{eqnarray}
\label{eq:expand}
   \ket{\Psi_j} &=&  \ket{\chi_j} \otimes \ket{\Phi_j} + \sum_{k \ne j} \frac{\bd_{jk} \cdot \hbP }{V_j - V_k}  \ket{\chi_k} \otimes \ket{\Phi_k}.
\end{eqnarray}
Nafie has shown that, if one follows such a procedure\cite{nafie:1983:jcp:el_momentum, coraline:2024:jcp:pssh_conserve}, one recovers a very desirable and intuitive expression for electronic momentum:
\begin{eqnarray}
\label{eq:elmoment}
    \bra{\Psi_j}\hbp \ket{\Psi_j} &=& m_e \frac{d \bra{\Phi_j} \hat{\bm x} \ket{\Phi_j}}{dt}.
\end{eqnarray}
In light of Nafie's work, one might wonder whether it is possible  to develop an electronic structure approach that directly produces electronic states $\ket{\tilde{\Phi}_j}$ satisfying $ \bra{\tilde{\Phi}_j} \hbp \ket{\tilde{\Phi}_j}  = m_e d\bra{\tilde{\Phi}_j}  \bm{\hat{x}} \ket{ \tilde{\Phi}_j}/dt$.   This challenge is one motivation for phase-space electronic structure theory described below, and one means of assessing the value of a beyond BO-approach  is the capacity to recover the intuitive expression in Eq. \ref{eq:elmoment} for single-state electronic momentum; the same criterion can also be used to judge recently developed exact-factorization approaches.\cite{abedi:2010:prl_exact_factorization, gross:2015:jcp:vcd_exact_factorization}

\subsubsection{Berry Phase, Berry Curvature and Quantum Geometry}
Whereas surface hopping techniques are commonly employed within chemistry to treat the ``well known''  static correlation nonadiabatic problems of Sec. \ref{sec-BO-fail1}, the ``lesser known'' dynamical correlation nonadiabatic problems described  above are often treated nowadays in the context of Berry curvature \cite{berry:1984:berryphase} and so-called ``quantum geometry''\cite{provost1980riemannian, kolodrubetz2017geometry}. To that end,  a word of context may be helpful for the reader as far as  understanding how the quantum geometry approach addresses the Born-Oppenheimer problems discussed above (and how such a quantum geometry approach differs from the phase-space electronic structure theory described below).  Note that quantum geometry techniques are far more common in physics than in chemistry. 

As discussed in Sec. \ref{sec-BO}, all of the errors in BO theory arise from the derivative couplings $\bd_{ij}^{I}$ -- which are very complicated objects.  In particular, in the context of photochemistry (Sec. \ref{sec-BO-fail1} above), we have focused on cases where off-diagonal elements of the derivative couplings are large and  drive transitions between electronic states (e.g. at curve crossings and conical intersections).  By contrast, for the problems of electronic momentum, the problematic elements of the derivative coupling tensor are usually small.  
Moreover, by time-reversibility $\bd_{kk}$ is zero for systems with an even number of electrons and a non-degenerate ground state. That being said, $\bd_{kk}$ does not vanish in the presence of an external magnetic field.  With this in mind, in order to incorporate electronic momentum into BO dynamics (say, along non-degenerate state $k$), the quantum geometry approach is to  modify Eq. \ref{eq:pdot:bo} above so to include the curl of the on-diagonal derivative coupling (which is also called the nuclear Berry curvature):
\begin{eqnarray}
\label{eq:berry:bomd}
    \dot{\bP} &=&  -\frac{\partial V_k}{\partial \bX}   + \bm{\Omega}_k \dot{\bm X},\\
\label{eq:berry:expand}
\bm{\Omega}_k &=&i\hbar\nabla \times  \bd_{kk} =-\hbar\sum_{j\neq k}\mbox{Im}\left( \bd_{kj} \times \bd_{jk} \right).
\end{eqnarray}
Note that, although $\bd_{kk}$ is not well-defined and depends on an arbitrary gauge (i.e. the phase of state $k$), $\nabla \times \bd_{kk}$  is well-defined. 

Intuitively, the Berry curvature keeps track of (partly) how the adiabatic states change with nuclear position, and thus perhaps not surprisingly, these matrix elements  can be used to correct BO theory for many purposes.
In particular, for our purposes,
including the Berry curvature in Eq. \ref{eq:berry:bomd} restores momentum conservation in the presence of spin-orbit coupling \cite{xuezhi:2023:total_ang_bomd} and pseudomomentum conservation in the presence of a $\bB$-field\cite{helgaker:2022:jcp_conservation_laws_magnetic_field}
More genearlly, very similar (strictly electronic) Berry curvature tensors are broadly used nowadays in solid-state physics to explain a host of other physical phenomena, especially electronic polarization and orbital magnetization\cite{vanderbiltbook}.


Notwithstanding all of the successes listed above, it must be emphasized that using the Berry curvature to ``fix up'' BO theory (in the spirit of Eq. \ref{eq:berry:bomd}) has severe limitations as far as incorporating nonadiabatic effects. First, as stated above, Eq. \ref{eq:berry:expand} is zero without the presence of some external factor to break time-reversibility, e.g. a $\bB$-field. Thus, Berry curvature approaches cannot be used {\em directly} to predict electronic momenta or VCD signals of closed shell molecules (though Ref. \citenum{resta:vcd:berry_curvature:2025} interprets Eq. \ref{eq:mfp} below as a mixed Berry curvature). Second and more generally, including Berry curvature   (as in Eq. \ref{eq:berry:bomd})
gives information about how nuclear dynamics are altered by electronic motion,  but does not yield  information about how the electronic system is affected by nuclear motion.  Third and perhaps most importantly, the Berry curvature in Eq. \ref{eq:berry:expand} cannot be applied and has little physical meaning in the case of degeneracy -- either orbital or spin degeneracy --  which is quite often the case for molecular chemistry (and always the case for open shell systems).  One can generalize the single-state Berry curvature in Eq. \ref{eq:berry:expand} to a multi-state (nonadiabatic or non-abelian) Berry curvature  (where the states occupy a subspace $\mathcal{S}$)  of the form\cite{wilczek1984appearance},
\begin{eqnarray}
\label{eq:nonad:berry_curvature}
\bm{\Omega}^{I\alpha, J\beta} &=&i\hbar\Big(\frac{\partial  {\hat{d}^{J\beta}}}{\partial X^{I\alpha}} - 
     \frac{\partial  {{\hat{d}}^{I\alpha}}}{\partial X^{J\beta}}\Big)+ [\hat{d}^{I\alpha},\hat{d}^{J\beta}]
     \\
\bm{\Omega}^{I\alpha, J\beta}_{kl}     & =& i\hbar\Big(\frac{\partial  {{d}^{J\beta}_{kl}}}{\partial X^{I\alpha}} - 
     \frac{\partial  {{{d}}^{I\alpha}_{kl}}}{\partial X^{J\beta}}\Big)+ \sum_{j\in \mathcal{S}}\Big({d}^{I\alpha}_{kj}{d}^{J\beta}_{jl} - {d}^{J\beta}_{kj}  {d}^{I\alpha}_{jl}\Big) \nonumber \\
     & = & -  \sum_{j\not\in \mathcal{S}}\Big({d}^{I\alpha}_{kj}{d}^{J\beta}_{jl} - {d}^{J\beta}_{kj}  {d}^{I\alpha}_{jl}\Big).
\end{eqnarray}
The resulting tensor does have meaning for Ehrenfest dynamics\cite{takatsuka:2005:jcp,krishna:2007:ehr_plus_berry} and is necessary to restore momentum conservation\cite{coraline:2024:jcp:Ehrenfest_conserve}. However, 
Ehrenfest mean-field dynamics have many limitations (e.g. they lack detailed balance\cite{tully:2005:detailedbalance,tully:2008:detailedbalance}) and thus cannot be used to generate equilibrium properties. In short, 
for many reasons, 
there is today a strong need to go beyond a Berry curvature (or quantum geometry) picture of coupled nuclear-electronic dynamics, an aim which we take up in the next section.

\section{Phase-Space Electronic Structure Theory}
Phase-space electronic structure theory aims to tackle many of the problems discussed above, offering a more accurate and balanced treatment of coupled nuclear-electronic motion that goes beyond what can be achieved with BO theory. 
Whereas phase space approaches have long been used to map quantum operators to classical operators\cite{kapral:1999:jcp} or to map electronic states to tractable harmonic spaces for quantum dynamical purposes\cite{miller:1997:mmst, stock_thoss:1997, liu2021unified}, 
the first {\em phase-space  electronic structure theory} was offered by Shenvi\cite{shenvi:2009:jcp_pssh} (and later investigated and expanded upon by
Barrera and co-workers\cite{barrera2025moving}).
Shenvi suggested that, if we consider $\bX$ and $\bP$  to be classical (rather than quantum) variables, one could go beyond the Born-Oppenheimer electronic Hamiltonian in Eq. \ref{eq:Hel} with an operator that accounted for nuclear motion:
\begin{eqnarray} \label{eq:HPShenvi}
   \hat H_{\rm Shenvi}(\bX,\bP) &=&    \sum_I \frac{{\left(\bm P_I - i\hbar \hat {\bm d}^I (\bX) \right)}^2}{2M_I} +\hat{H}_{\rm el}({\bm X}). 
\end{eqnarray}
Eq. \ref{eq:HPShenvi} is identical to Eq. \ref{eq:Hadfull}, but we replace $(\hbX,\hbP)$ with $(\bX,\bP)$.  
Now, in his original paper,\cite{shenvi:2009:jcp_pssh} 
Shenvi was  looking to treat the static correlation problem, as his goal was to create a new phase-space surface hopping algorithm (that went beyond traditional surface hopping\cite{truhlar:review:surfacehop}). 
However, while Shenvi's approach  had some clear successes in the adiabatic limit\cite{shenvi:2009:jcp_pssh,izmaylov:2016:jpc_dboc_pssh}, it also had a few failures\cite{izmaylov:2016:jpc_dboc_pssh}. In particular, we now understand that whereas a phase-space approach can be used to improve upon 
{\em dynamic} correlation,  Eq. \ref{eq:HPShenvi} cannot fix up the  
{\em static} correlation. In particular, if one  inserts the the derivative couplings in Eq.  \ref{eq:HPShenvi},
the potential energy surfaces become unstable near crossings and intersections (where $\bd$ gets very large [see Eq. \ref{eq:derv_coupl_hellman_feynman}]).  Note also the phase-space potential energy surfaces are not well-defined if the BO electronic states are degenerate (for example, for spin doublets) as the derivative couplings are not well-defined. Finally, the entire procedure is expensive and bulky; in principle, one must diagonalize the BO electronic Hamiltonian in Eq. \ref{eq:Hel}, generate standard adiabatic electronic states, evaluate the derivative couplings between those state, then rebuild the Shenvi Hamiltonian in Eq. \ref{eq:HPShenvi}, and then rediagonalize again. 
For all of these reasons, as far as we are aware,  there have never been any {\em ab initio} simulations with Eq. \ref{eq:HPShenvi} heretofore.

For all of these reasons, the more recent push into phase-space electronic structure theory 
does not focus on the static correlation problem but rather on the dynamical correlation problem.  In particular,  one can show that, in order to restore a nonzero electronic momentum beyond BO theory in a meaningful and tractable fashion, one needs only to incorporate some element of the derivative couplings from Eq. \ref{eq:Hadfull} and one can work with an approximate operator $\hbGamma_I$, 
\begin{eqnarray} \label{eq:HPS}
   \hat H_{\rm PS}(\bX,\bP) &\equiv&    \sum_I \frac{{\left(\bm P_I - i\hbar \hat {\bm \Gamma}_I (\bX) \right)}^2}{2M_I} +\hat{H}_{\rm el}({\bm X}). 
\end{eqnarray}
Here,  if $\hbGamma_I$ is set to zero, the phase-space electronic Hamiltonian $\hH_{\rm PS}$ becomes equivalent to the Born-Oppenheimer electronic Hamiltonian (up to a constant factor of kinetic energy), $\bP^2/(2M) + \hH_{\rm el}$.
In a moment, we will hypothesize a meaningful form for $\hbGamma$, but before doing so, it should be clear that, if we seek an approximate operator that mimics the derivative coupling, that operator ($\hbGamma$) must satisfy  equations corresponding to Eqs. \ref{eq:dconstrain1a}-\ref{eq:dconstrain4a} above for $\bd$:

\begin{eqnarray}
    -i\hbar\sum_{I}\hbm{\Gamma}_{I} + \hat{\bm p} &=& 0,\label{eq:constrain1}  \\
    \Big[-i\hbar\sum_{J}\pp{}{\bm{X}_J} + \hat{\bm p}, \hbm{\Gamma}_I\Big] &=& 0,\label{eq:constrain2}\\
    -i\hbar\sum_{I}{\bm X}_{I} \times \hat{\bm \Gamma}_{I} + \hat{\bm l} + \hat{\bm s} &=& 0,\label{eq:constrain3}\\
     \Big[-i\hbar\sum_{J}\left(\bm{X}_J \times\pp{}{\bm{X}_J}\right)_{\gamma} + \hat{l}_{\gamma} + \hat{s}_{\gamma}, \hat{\Gamma}_{I \delta}\Big] &
     =& i\hbar \sum_{\alpha} \epsilon_{\alpha \gamma \delta} \hat{\Gamma}_{I \alpha}. \nonumber \\
     \label{eq:constrain4} 
\end{eqnarray}

Indeed, let us imagine that one builds and diagonalizes a phase-space electronic Hamiltonian of the form in Eq. \ref{eq:HPS} with a $\hbGamma$ operator that satisfies Eqs. \ref{eq:constrain1}-\ref{eq:constrain4}, 
\begin{eqnarray}
\label{eq:diagHPS}
    \hH_{\rm PS}(\bX,\bP) \ket{\Phi^{\rm PS}_j (\bX,\bP)} = 
        V_j^{\rm PS}(\bX,\bP) \ket{\Phi^{\rm PS}_j (\bX,\bP)}.
\end{eqnarray}
In such a case, if one runs  classical dynamics along a resulting eigensurface $k$, 
\begin{eqnarray}
    \dot{\bX} = \frac{\partial  V^{\rm PS}_k}{\partial \bP} \\   
    \dot{\bP} = -\frac{\partial V^{\rm PS}_k}{\partial \bX} 
\end{eqnarray}
the dynamics are guaranteed to conserve the total linear and  angular momentum\cite{coraline:basisfree:2025} (in addition to the energy), while also imposing a nonzero electronic momentum (see Sec. \ref{sec-vcd} below).

\subsection{Form of the {$\mathbf{\hat{\Gamma}}$} operator}
As far as implementing a phase-space electronic Hamiltonian, all that remains is to specify the  exact form of $\hbGamma$. We postulate that we can write $\hbGamma$ as the sum of two contributions:
\begin{eqnarray} 
\label{eq:gamma_sum}
    \hat{\bm \Gamma}_A' = 
        \hat{\bm \Gamma}_A'
        +     \hat{\bm \Gamma}_A'',
\end{eqnarray} 
The first contribution ($\hat{\bm \Gamma}_A'$) captures how electrons are dragged along by nuclei (which is equivalent to expressing the electronic DOF relative to the nuclear center of mass, as in Sec. \ref{sec-hydrogen} below). 
The second contribution ($\hat{\bm \Gamma}_A'$) captures how electrons are rotated together with nuclei (which is equivalent to expressing the electronic DOF relative to the nuclear orientation). $\hat{\bm \Gamma}_A'$
is necessary for conserving the total linear momentum; $\hat{\bm \Gamma}_A''$
is necessary for conserving the total angular momentum.
To that end, we will now postulate a given form for $\hbGamma'$ and $\hbGamma''$; we will motivate  these choices  shortly thereafter in Sec. \ref{sec-sigma-inf}.

\subsubsection{Form of $\hat {\bm \Gamma}_A'$}

For the construction of $\hat{\bm \Gamma'}$, we will use a simple version of electron translation factors (ETFs) from Ref. \citenum{coraline:basisfree:2025}:
\begin{eqnarray} \label{eq:gamma1}
    \hat{\bm \Gamma}_A' = \frac 1 {2i\hbar} \left( \hat{\theta}_A \hat{\bm p} + \hat{\bm p}   \hat{\theta}_A\right).
\end{eqnarray}
Here, $\hat \theta_A$ is a so-called local partition of unity function for which  $\hat \theta_A(\bx)$ is near unity when $\bx$ is near $\bX_A$,  $\hat \theta_A(\bx)$ is zero when  $\bx$ is far from $\bX_A$, and
\begin{eqnarray}
  \sum_A \hat{\theta}_A(\bx) = 1.
\end{eqnarray}

We have chosen to define $\hat \theta_A$ as combinations of Gaussian functions localized on different atoms: 
\begin{eqnarray}
\label{eq:theta}
    \hat \theta_A(\hat{\bm x}) = \frac{Q_Ae^{-|\hat{\bm x}-\bm{X}_A|^2/\sigma^2}}{\sum_B Q_Be^{-|\hat{\bm x}-\bm{X}_B|^2/\sigma^2}}.
\end{eqnarray}
In Eq.~\ref{eq:theta}, we utilize  Gaussian functions with width $\sigma$, centered around $\bm X_A$ and weighted by the charge on atom $A$, $Q_A$.  
The physical meaning of this spatial partition function $\hat \theta_A$ is  clear; $\hat \theta_A$ measures the extent to which the electrons are influenced by nucleus $A$, i.e., electrons are more strongly affected by nucleus $A$'s motion when they are closer to $A$ and when $A$ has more charge; the parameter $\sigma_A$ determines the precise measure of locality. The parameter $\sigma$ defines the length or width over which the partitioning scheme changes between ``atomic sectors''.

\subsubsection{Form of $\hat {\bm \Gamma}_A''$}
For the rotational component $\hat {\bm \Gamma}_A''$, in analogy to the case above, we will employ the basis-free electronic rotation factors  (ERFs) from Ref. \citenum{tian:2024:jcp:erf}: 
\begin{eqnarray}
    \hat{\bm \Gamma}_A'' &=& \sum_B\zeta_{AB}\left(\bm{X}_A -\bm{X}^0_{B}\right)\times \left(\bm{K}_B^{-1}\hat{\bm J}_B\right),\label{eq:erf_final}
\end{eqnarray}
where 
\begin{widetext}
\begin{eqnarray}
    \hat{\bm J}_B &=& \frac{1}{2i\hbar}\left((\hat{\bm x}-\bm{X}_B)\times \hat \theta_B(\hat{\bm x} )\hat{\bm p} +(\hat{\bm x} -\bm{X}_B)\times\hat{\bm p} \hat \theta_B(\hat{\bm x})+2\hat{\bm s} \hat{\theta}_B(\hat{\bm x})\right), \\
    \bm{X}_{B}^0 &=& \frac{\sum_A\zeta_{AB}\bm{X}_A}{\sum_A\zeta_{AB}},\\
    \label{eq:KB}
    \bm{K}_B &=& \sum_A\zeta_{AB}\left(\bm{X}_A\bm{X}_A^\top-\bm{X}_B^0\bm{X}_B^{0\top}-(\bm{X}_A^\top\bm{X}_A-\bm{X}_B^{0\top}\bm{X}_B^0)\mathcal{I}_3\right),
\end{eqnarray}    
\end{widetext}
and the  localization function $\zeta_{AB}$ is
\begin{eqnarray}
\label{eq:zeta}
    \zeta_{AB} &=M_A e^{-|\bm{X}_A-\bm{X}_B|^2/\beta_{AB}^2}.
\end{eqnarray} 

Intuitively, with this prescription, we are considering how all atoms near atom $A$ are positioned around said atom, and then we use this instantaneous information about the orientation of the local nuclear frame to concurrently rotate the local electronic frame.
Here, $\beta_{AB}$ is the length scale over which one expects the rotations of nearby nuclei to couple to the angular momentum of the electrons. 

\subsubsection{The Limit $\sigma \rightarrow \infty$}\label{sec-sigma-inf}
It is straightforward to verify that the $\hat {\bm \Gamma}$ term defined above satisfies the four constraints in Eqs.~\ref{eq:constrain1}-\ref{eq:constrain4}\cite{coraline:basisfree:2025}.
Altogether, Eqs.~\ref{eq:gamma_sum}-\ref{eq:zeta} constitute a complete, practical, self-contained expression for $\hbGamma$ from which one can build a phase-space Hamiltonian $\hat H_{\rm PS}$. 
Admittedly, this choice for $\hbGamma$ is not unique. For instance, the charge factor $Q_A$ present in Eq. \ref{eq:theta} or mass factor $M_A$ present in Eq. \ref{eq:zeta} can be replaced by any other weighting function without violating  Eqs.~\ref{eq:constrain1}-\ref{eq:constrain4}\cite{coraline:basisfree:2025}. That being said,
Eqs.~\ref{eq:gamma_sum}-\ref{eq:zeta} above have one strong endorsement. Namely, in the limit that we ignore all locality and set $\sigma \rightarrow \infty$ and if we assume that the charges and masses are proportional to each other, so that $Q_A/M_A$ is a constant, one can show that the resulting phase-space electronic Hamiltonian is equivalent to\cite{coraline:basisfree:2025}:
\begin{widetext}
 
\begin{eqnarray}
   \hat H_{\rm PS}(\bX,\bP) &=& \sum_A \frac{ \left( \bm P_A - \frac{M_A}{M_{\rm tot}} (\sum_B \bm P_B) \right) ^2}{2M_A} 
+ \frac{(\sum_A  \bm P_A -\hbm{p})^2}{2M_{\rm tot}}\nonumber  \\
& & - \left(\sum_A (\bX_A-\bX_{\rm NCM}) \times \bP_A\right) \cdot \bI_{\rm NCM}^{-1} \cdot \left( \hbm{l}_{\rm NCM} +\hbm{s} \right)  \nonumber   \\
& & +  \frac{1}{2} \left( \hbm{l}_{\rm NCM} +\hbm{s} \right) \cdot  \bI_{\rm NCM}^{-1} \cdot \left( \hbm{l}_{\rm NCM} +\hbm{s} \right).
      \label{eq:HPS_sigma_inf}
\end{eqnarray}
   
\end{widetext}

Here, $\bX_{\rm NCM}$ is the nuclear center of mass, $\bI_{\rm NCM}$ is the total nuclear moment of inertia matrix relative to the nuclear center of mass, and $\hbm{l}_{\rm NCM}$ is the electronic angular momentum relative to the nuclear center of mass.
Thus, in the limit $\sigma \rightarrow \infty$, four terms appear: (i) the kinetic energy of the atoms relative to the center of nuclear mass, (ii) the total  kinetic energy of the nuclear center of mass (recognizing that $\sum_A \bm P_A$  now represents the total nuclear plus electronic momentum, from which one must subtract the electronic momentum)\cite{littlejohn:2023:jcp:angmom}, (iii) a coriolis potential from aligning the electronic frame of the electrons to the nuclei \cite{littlejohn:flynn:1991:pra:coriolis}, and (iv) a centrifugal potential that captures the inertia of electrons in a rotating frame\cite{aucar:2011:coriolis}.
This behavior is exactly what we would hope from a rigid molecular ensemble interacting with a set of mobile electrons. Thus, by reducing $\sigma$ from infinity to a finite value, we are merely allowing for electrons to revolve and equilibrate around smaller domains (made up of one or a few nuclei) and enforcing size consistency (i.e. the notion that all dynamics must be local and systems separated far apart do not interact); by  separating out the local center of mass and local frame of orientation, we can identify new physical effects along a single electronic surface.

\subsection{Advantages and Disadvantages of Phase-Space Electronic Structure }
In Sec. \ref{sec-ps-success} below, we will highlight in detail several successes of the phase-space electronic structure theory. That being said, even without thinking too hard, it should be clear that phase-space methods will be most readily applied in a semiclassical context. Indeed, if $\bX$ and $\bP$ were quantum operators (i.e. $\hbX$ and $\hbP$), then diagonalizing Eq. \ref{eq:HPS} would be no simpler than diagonalizing the full Schrodinger equation (Eq. \ref{eq:Htot}). That being said, if $\bX$ and $\bP$ are considered to be classical, then diagonalizing Eq. \ref{eq:HPS} is quite similar in cost to diagonalizing the standard BO electronic Hamiltonian (Eq. \ref{eq:Hel}).  After all, the major addition ($\hbGamma$) is only a one-electron term which carries very little added cost. The only major difference between a phase-space versus a standard BO electronic structure calculations is that now the Schrodinger equation becomes complex-valued (with real valued energies but complex-valued wavefunctions)--which does necessitate  rewriting major sections of existing electronic structure code.

Now it is important to note that many chemistry and physics questions can be answered with semiclassical (as opposed to quantum) nuclei\cite{heller:1981:acr}. First and foremost, equilibrium stationary structures do not need quantum nuclei, and so PS approaches offer us the capacity to explore new equilibrium geometries -- both as a function of $\bX$ {\em and} $\bP$. Indeed, in Sec. \ref{sec-spin} below, we will discuss how, when spin degrees of freedom are present, new stationary points can emerge with major consequences for magnetic dynamics. Second, many nonadiabatic  problems can be answered with semiclassical dynamics, especially Ehrenfest\cite{doltsinis:2002:review} and surface hopping\cite{barbatti:2011:review} dynamics; in fact, it is quite common today to run large semiclassical nonadiabatic simulations to simulate photoinduced processes,\cite{tretiak:2014:acr} where electronic transitions occur routinely and energy dissipation from the electronic degrees of freedom is coupled with excitation of nuclear degrees of freedom. Within the context of phase-space surface hopping\cite{bian:2024:pssh_translations_rotations}, however, we now will be able to also monitor the exchange of linear and angular momentum between nuclei and electrons and spin (all while conserving the total)\cite{gross:2022:prl:angmom}.
As such, as discussed in Sec. \ref{sec-app} below, we should also have the capacity to simulate many new phenomena  in  new ways.

So far, we have discussed the strong advantages of a phase-space electronic structure approach. That being said, one can ask: what are the disadvantages?
Perhaps the strongest disadvantage of a phase-space approach is that, in order to go beyond a semiclassical ansatz and extract  quantum nuclear information, we must work with Wigner transforms (Eq. \ref{eq:wigner}) and Weyl transforms (Eq. \ref{eq:weyl}), here written for some  operator $\hat{O}$\cite{wigner1932, balazs1984}:
\begin{eqnarray} 
    \hat O_W(\bX,\bP) &=& 
    \int d\bm X' \bra{\bm X + \frac {\bm X'} 2 } \hat O \ket{\bm X - \frac {\bm X'} 2 } e^{-\frac i \hbar \bm X' \cdot \bm P}, \nonumber \\ \label{eq:wigner} \\ 
    \bra{\bm X} \hat{O} \ket{\bm X'}  
    & = & \int \frac {d\bm P} {2\pi\hbar} e^{\frac i\hbar \bm P \cdot (\bm X - \bm X')} \hat O_W\left(\frac {\bm X + \bm X'} 2, \bm P\right). \nonumber \\
        \label{eq:weyl}
\end{eqnarray}
In other words, mathematically, we must consider the parameters $\bX,\bP$  in Eq. \ref{eq:HPS} to be Wigner symbols, and if we seek quantum matrix elements, we must necessarily apply a Weyl transform to requantize. 

This need for Wigner-Weyl transforms has two consequences. First, if we seek to calculate the exact solution to a quantum problem using a set of phase-space electronic basis functions, the resulting Hamiltonian is  complicated because we must use several Moyal brackets and other  tricks of Wigner-Weyl transforms\cite{case:2008:wigner_review}. Put bluntly, one can view Eq. \ref{eq:Hadfull} above as an expansion in $\hbar,$ for which the total Hamiltonian terminates at second order in the BO framework,
\begin{eqnarray}
    \hH = \hH_{\rm BO}^{(0)} + \hH_{\rm BO}^{(1)} + \hH_{\rm BO}^{(2)}.
\end{eqnarray}
By contrast,   in a phase-space framework,  the total Hamiltonian becomes an infinite, perturbative sum in $\hbar$ that operates locally but never terminates\cite{xinchun:2025:wigner_one_state}. We note that such an infinite summation was developed by Littlejohn and Flynn (LF)\cite{littlejohn:flynn:1991:pra:coriolis} and Teufel\cite{matyus2019} (following earlier work by Blount\cite{blount:1962:partwig}) in the context of a BO zeroth order calculation,
\begin{eqnarray}
\label{eq:hlf:expand}
       \hH = 
    \hH_{\rm BO}^{(0)} +
        \hH_{\rm LF}^{(1)} +
            \hH_{\rm LF}^{(2)} +
                \hH_{\rm LF}^{(3)} + \ldots.
\end{eqnarray}
A similar exact expansion can also be developed for a phase-space zeroth order calculation\cite{xinchun:2025:wigner_one_state}:
\begin{eqnarray}
\label{eq:hps:expand}
       \hH = 
    \hH_{\rm PS}^{(0)} +
        \hH_{\rm PS}^{(1)} +
            \hH_{\rm PS}^{(2)} +
                \hH_{\rm PS}^{(3)} + \ldots.
\end{eqnarray}
In our limited experience\cite{xinchun:2025:wigner_one_state} with model problems, including $\hH_{\rm PS}^{(1)}$ can strongly improve vibrational energies  for eigenstates that do not visit regions of configuration space near avoided crossings or conical intersections. That being said, including higher order corrections becomes expensive.
 
Second, working with Wigner transforms becomes prohibitively difficult in many dimensions.  Thus, if we seek  exact quantum nuclear information beyond a harmonic limit, phase-space methods will be challenging (though admittedly, BO-based methods are also challenging as quantum mechanics is exponentially complex).  That being said, the approach can useful for one or two dimensions (see Sec. \ref{sec-xuezhi-vib} below for one example).

Apart from the above two disadvantages, there is one other slightly uncomfortable feature of phase-space methods. Namely, the method requires the definition of a $\hbGamma$ operator, which is not uniquely defined (and thus can depend on the chemist running the calculation). Moreover, even if one is satisfied with the class of operators we have proposed above (in Eqs. \ref{eq:gamma_sum}-\ref{eq:zeta}), it is worth noting that these operators still require a width parameter $\sigma$ (in Eq. \ref{eq:theta}) and a rotational cutoff $\beta_{AB}$ (in Eq. \ref{eq:zeta}).  Thus, no matter how one slices it, phase-space methods do require some degree of parameterization, and in principle, poor parameters could lead to suboptimal calculations. 
Despite these disadvantages, it is  worth emphasizing that (i) with proper benchmarking, good parameters can be  found, such that one can extract electronic momenta that are better than zero (see Fig. \ref{fig:elmom} below);  (ii) in principle, following Eq.  \ref{eq:hps:expand}, one can always go to higher order if one is not convinced that the zeroth order is accurate (just as one can also improve accuracy by including some nonadiabatic effects on top of a BO ground-state calculation). Thus, even though a phase-space framework requires paramterization, the method can be extended to an exact {\em ab initio} Hamiltonian. Moreover,
it is worth noting that uniqueness is not the ultimate goal of science.  Rather, improved accuracy and the capacity to discover new physics is a higher prize, according to which it is always best to choose the partition that best minimizes the overall error -- which is exactly the motivation for the phase-space electronic Hamiltonian approach.

With this background in mind, let us now discuss the key successes of the phase-space approach as reported thus far.

\section{Phase-Space Successes  Without Spin}
\label{sec-ps-success}
\subsection{An exact treatment of the hydrogen atom}\label{sec-hydrogen}
One of the clearest failures of the standard BO approximation is the inability to recover the exact electronic eigenergies for the hydrogen atom. As found in any text book\cite{griffiths:quantum_textbook:2018}, the exact energies of the hydrogen atom are:
\begin{eqnarray}\label{eq:Eexact}
    E_{n}^{\rm exact}(\bm P) = \frac{\bm P^2}{2(M_H + m_{\rm e})} -\frac{\mu e^4}{8 \epsilon_0^2 h^2} \frac{1}{n^2}.
\end{eqnarray}
where $\mu$ is the reduced mass of the electron, $1/\mu = 1/m_e + 1/M_H$.  That being said, in order to recover this answer, one must separate out the center of mass from the relative coordinate. This separation is not part of the standard BO approach, which instead recovers energies of the form 
\begin{eqnarray} \label{eq:Ecn}
    E_{n}^{\rm BO}(\bm P) = \frac{\bm P^2}{2M_H } -\frac{m_{\rm e} e^4}{8 \epsilon_0^2 h^2} \frac{1}{n^2}.
\end{eqnarray}
with the raw electronic mass $m_e$ instead of the reduced mass $\mu$.

Now, within the phase-space electronic structure approach, the phase-space electronic Hamiltonian is:
\begin{eqnarray} \label{eq:PSHatom}
    \hat{H}_{\rm PS} 
    & = & \frac{({\bm P} - \hbp)^2}{2M_H} +\hat{H}_{\rm el}, 
\end{eqnarray}
For this model problem, with only one nucleus, the $\hbGamma$ operator is particularly simple, $i \hbar \hbGamma = \hbp$. 
If one diagonalizes Eq. \ref{eq:PSHatom}, it is straightforward to show that one recovers the exact hydrogenic eigenspectrum in Eq. \ref{eq:Eexact}. In other words, the phase-space electronic Hamiltonian is exact for the hydrogen atom\cite{bian:2025:jctc:wigner_vibrations}. Moreover, not coincidentally, phase-space electronic structure theory recovers 
the exact electronic linear momentum for a translating system.

\subsection{Vibrational Energies of Model Systems}
\label{sec-xuezhi-vib}

One of the most interesting features of phase-space electronic Hamiltonians is that, even though the method is best suited for semiclassical approximations, as discussed above, one can in fact extract quantum energies as well -- if one is prepared to perform a Weyl transform. To see how such a procedure can be performed in practice, imagine that we diagonalize Eq. \ref{eq:diagHPS} to generate $V_j^{\rm PS}(\bX,\bP)$. 
Now, fix $j=0$ so that we are working on the ground state.  To zeroth order, we can construct a fully quantum vibrational Hamiltonian by simply requantizing the operator through Eq. \ref{eq:weyl}:
\begin{eqnarray}
    \bra{\bm X} \hat{V}^{\rm PS}_0 \ket{\bm X'}  
    & = & \int \frac {d\bm P} {2\pi\hbar} e^{\frac i\hbar \bm P \cdot (\bm X - \bm X')} V^{\rm PS}_0\left(\frac {\bm X + \bm X'} 2, \bm P\right) \nonumber \\
\end{eqnarray}

\begin{figure*}
    \centering \includegraphics[width=\linewidth]{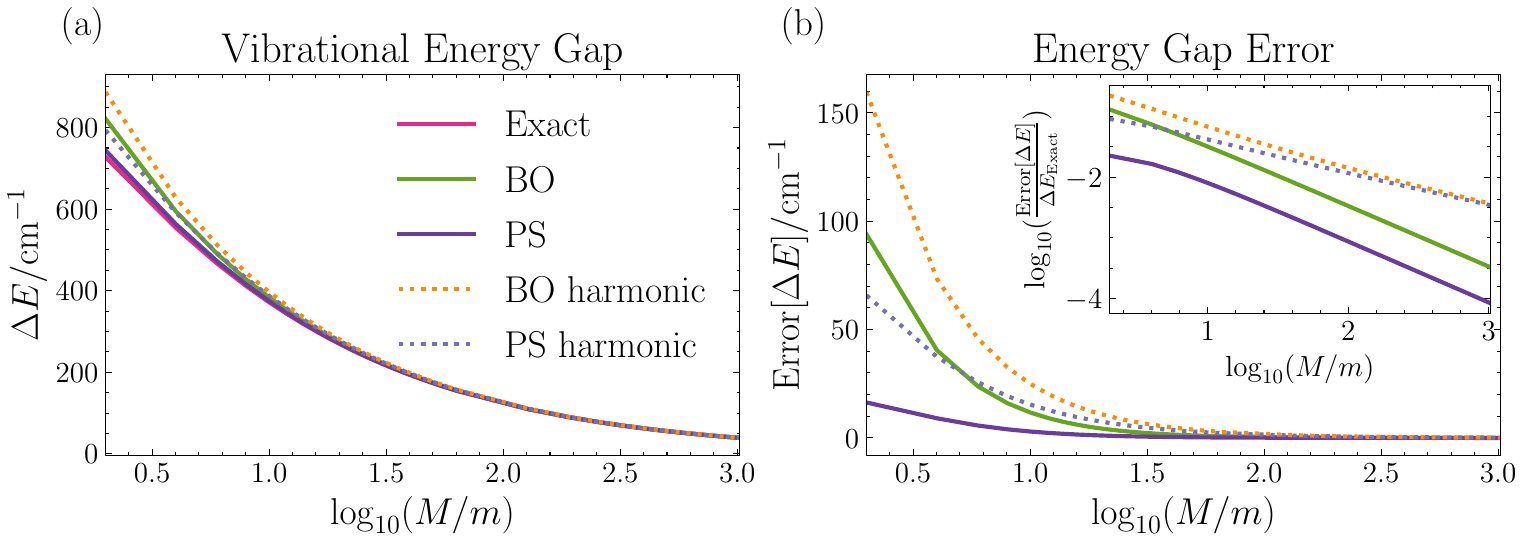}
    \caption{(a) The vibrational energy gap ($\Delta E = E_2 - E_1$) according to exact dynamics, BO dynamics (exact and harmonic), PS dynamics (exact and harmonic) for the proton transfer Hamiltonian designed in Refs. \cite{borgis:model:xuezhi:gross:wigner:cpl:2006,gross:2017:prx:born_oppenheimer_mass} and studied in Ref. \cite{bian:2025:jctc:wigner_vibrations} (b) The error in the energy gap relative to exact as well.  Note that PS theory strongly outperforms BO methods, especially for reasonably light nuclear masses.  }
    \label{fig:xuezhi}
\end{figure*}

In Fig. \ref{fig:xuezhi}, we explore a model Hamiltonian studied by Borgis {\em et al}\cite{borgis:model:xuezhi:gross:wigner:cpl:2006} and later Gross {\em et al}\cite{gross:2017:prx:born_oppenheimer_mass} which was designed originally to mimic proton transfer with one ``light'' particle and two ``heavy'' particles.  
%
For such Hamiltonian, we evaluated the two lowest quantum energies (they are effectively vibrations) as predicted by BO theory and PS theory (all relative to the exact answer). For BO theory, we diagonalize $\frac{\hbP^2}{2M} + \hV^{\rm BO}_0$ (see Eq.  \ref{eq:Hel_diag}); for PS theory, we diagonalize $\bra{\bm X} \hat{V}^{\rm PS}_0 \ket{\bm X'}$.  In Fig. \ref{fig:xuezhi}, we plot results for the $E_1 - E_0$ gap for approximate methods vs exact diagonlization as a function of the nuclear mass. 
The PS method always outperforms BO theory, and this improvement becomes quite large as the mass of the light particle becomes heavier and heavier. For more details, see Ref. \citenum{bian:2025:jctc:wigner_vibrations}.  Fig. \ref{fig:xuezhi} is another strong vindication of using a PS approach. 

\subsection{{\em Ab Initio} \label{sec-vcd} Calculations of Vibrational Circular Dichroism}
As discussed above (see Sec. \ref{sec-BO-fail2}),  vibrational circular dichroism is one experiment where one can directly test the validity of a phase-space electronic structure Hamiltonian.
On the one hand, BO theory can certainly predict a nonzero electric dipole between two ground state vibrational states (say 1 and 2); as is well known in infrared (IR) spectroscopy, a good approximation is $\bmu_{12} = {\partial {\bmu_g}}/{\partial \bX} \cdot \bX_{12}$. That being said, just as BO theory  cannot directly predict a ground state electronic momentum, BO theory cannot predict a ground state magnetic electronic  dipole moment  ($\bmm_{12} =0$); therefore BO theory cannot directly make a meaningful prediction of VCD spectra.  At present, the most popular method to extract a VCD rotatory strength is magnetic field perturbation (MFP) theory\cite{stephens:1985:jpcc_vcd, stephens:1987:gauge:vcd,BUCKINGHAM:1987:vcd}, where one asserts that the magnetic dipole moment (say, between states 1 and 2) can be calculated by double perturbation theory as 
\begin{eqnarray}
\label{eq:mfp}
\frac{\partial \bmm}{\partial \bP_A} = \frac{2i\hbar}{M_A} \bra{\frac{\partial \Phi}{\partial \bB}} \ket{\frac{\partial \Phi}{\partial \bX_A}}
\end{eqnarray}
While the  results using Eq. \ref{eq:mfp} are often good, there is something quite unsatisfying about this approach. After all,
for the electric dipole moment, one requires only a simple derivative or perturbation; why does the magnetic dipole moment require a double perturbation? More generally, VCD is a linear spectroscopy and all other linear spectroscopies (e.g. IR) require only a single perturbation. Why should VCD require a double perturbation theory? Lastly, whereas the IR spectrum can be modeled by fourier transforming the dipole-dipole time correlation function  $ \langle \bmu (0) \bmu (t) \rangle$ , should it not be possible to evalutate VCD directly through the time correlation  function $ \langle \bmu (0) \bmm(t) \rangle$?

\begin{figure*}
    \centering \includegraphics[width=\linewidth]{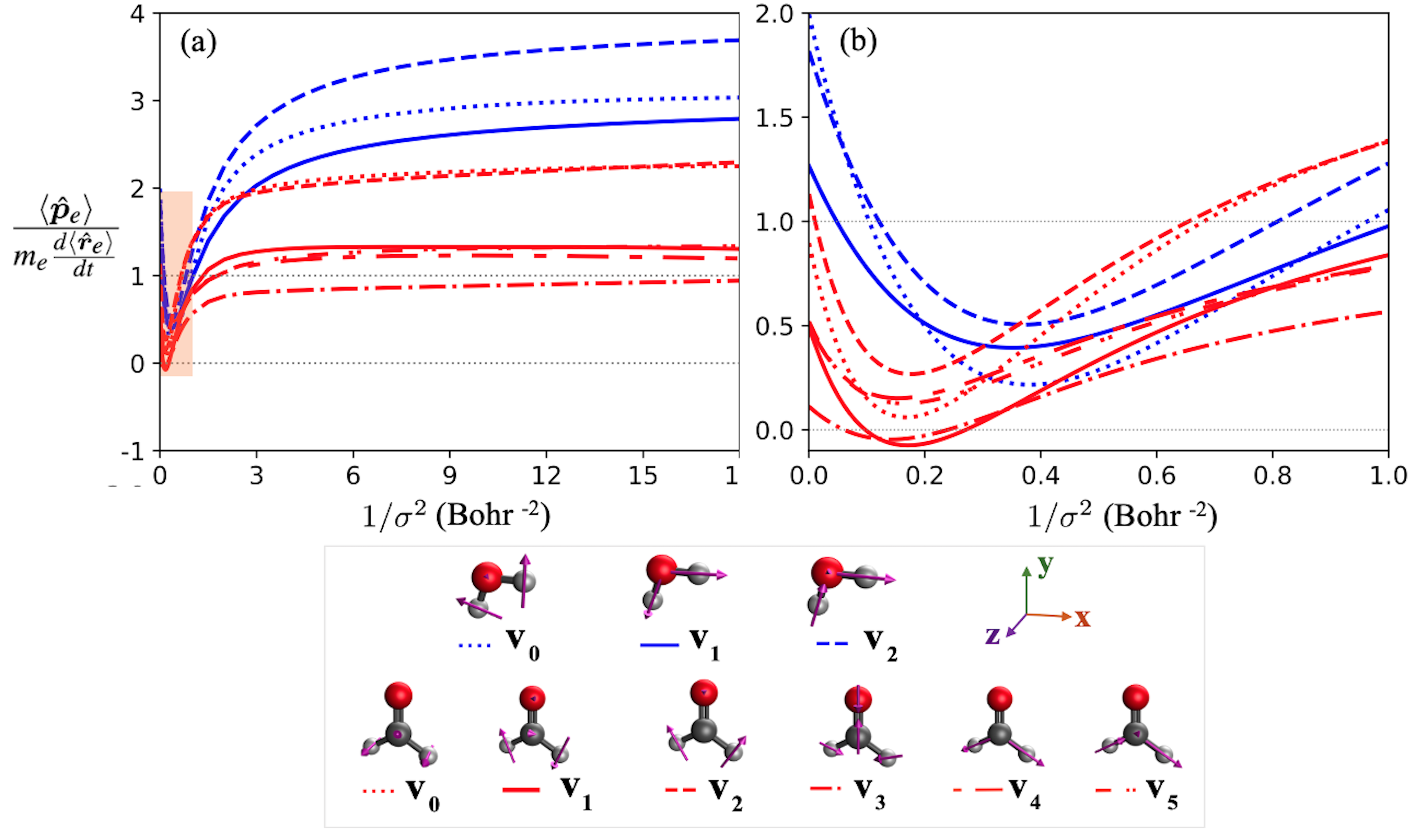}
    \caption{We investigate the linear electronic momentum for the vibrational modes of water (H$_2$O) and formaldehyde (CH$_2$O) using phase-space methods (and $\theta$  from Eq. \ref{eq:theta}). Here, in (a) we plot the ratio of $\bra{\Phi_{\rm PS}} \hbp \ket{\Phi_{\rm PS}}$ divided $m_e d\bra{\Phi_{\rm BO}} \hbx \ket{\Phi_{\rm BO}}/dt $  as a function of $\sigma$. The orange shaded area at large $\sigma$ values in (a) is enlarged and shown in (b). Note that, if $\sigma$ is small enough,  we recover the correct sign and the right order of magnitude, highlighting the power of a phase-space approach. Data from Ref. \citenum{coraline:basisfree:2025}}
    \label{fig:elmom}
\end{figure*}

One of the major  successes of phase-space electronic structure theory is the capacity to calculate VCD directly and avoid the uncomfortable questions above. In particular, within phase-space electronic structure theory, one can calculate a non-vanishing electronic momentum and $\partial \bmm/\partial \bP$ directly. 
To that end, in Fig. \ref{fig:elmom}, we investigate the electronic linear momentum as calculated for the normal modes of water (H$_2$O) and formaldehyde (CH$_2$O).  We evaluate the ratio of $\bra{\Phi_{\rm PS}}\hbp \ket{\Phi_{\rm PS}}$ divided $m_e d\bra{\Phi_{\rm BO}}\hbx \ket{\Phi_{\rm BO}}/dt $ as a function of the width parameter $\sigma$ from Eq. \ref{eq:theta}.  In principle, one would like to find an answer of unity if one were calculating the exact electronic momentum \cite{nafie:1983:jcp:el_momentum,coraline:2024:jcp:pssh_conserve} -- but this result follows from including both static and dynamic correlation (and we have already given up on static correlation).  Nevertheless, provided $\sigma$ is small enough,  we always recover the correct sign and the results are of the correct order of magnitude.

\begin{figure}
    \centering \includegraphics[width=\linewidth]{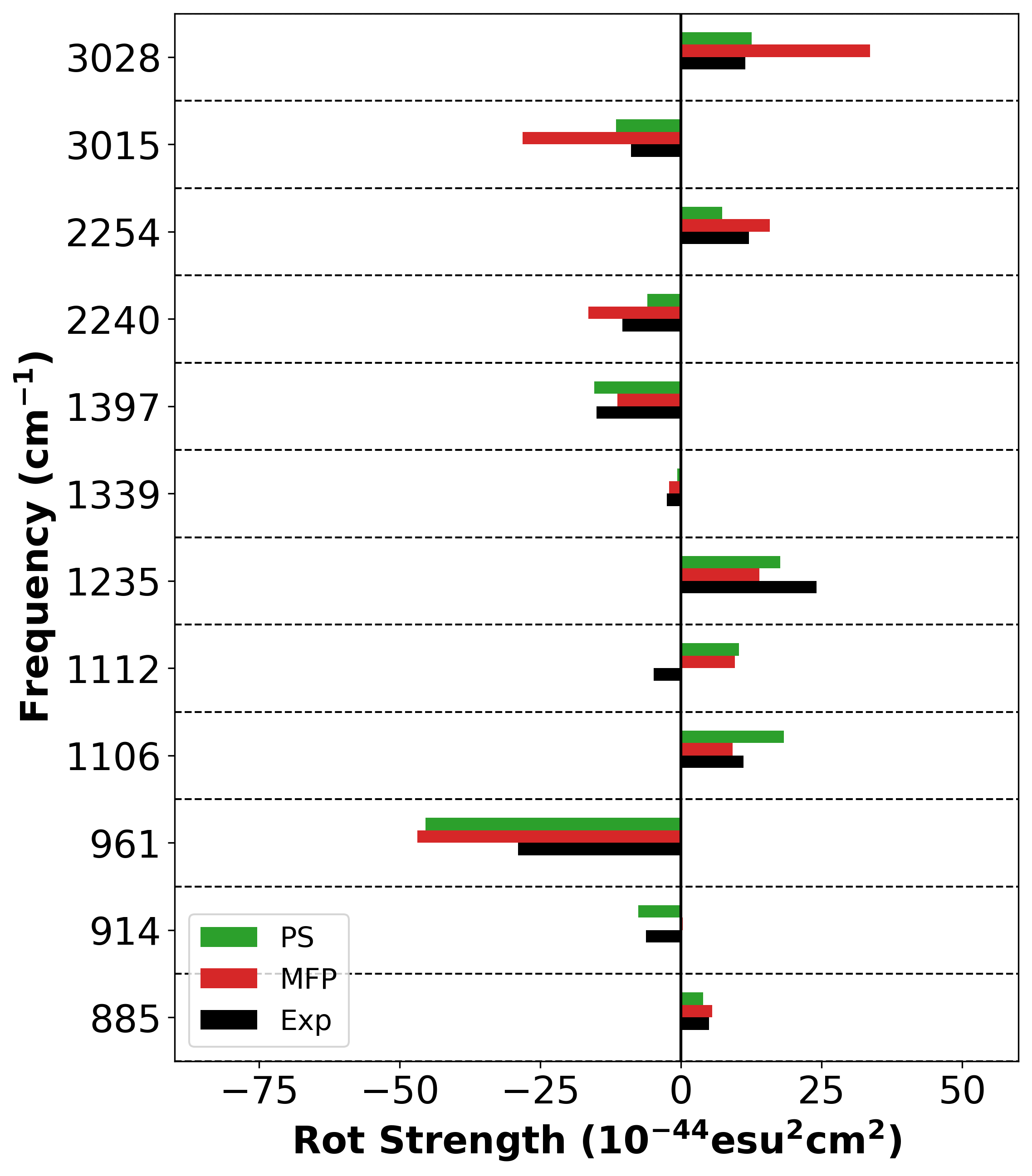}
    \caption{For the molecule (2S,3S)-oxirane-d2, with harmonic HF/cc-pvqz normal modes, we plot the experimental VCD signals (in solution)\cite{freedman1991vibrational} vs MFP predicted VCD spectra\cite{stephens:1985:jpcc_vcd,stephens:1994:cpl:exp_vcd_cycp} vs phase-spaced predicted VCD spectra. Not that a phase-space electronic Hamiltonian matches experimental data quite well, further validating our approach.  
    PS data adapted and updated from Ref. \citenum{zhentao:2024:jcp:vcd_basis_free} from a GHF calculation with the cc-pvtz basis set.}  
    \label{figvcd}
\end{figure}

Next, let us turn to VCD experimental data.  In Fig. \ref{figvcd}, we plot VCD results comparing phase-space theory to experiment for the molecule (2S,3S)-oxirane-d\cite{freedman1991vibrational}; we also include MFP results (Eq. \ref{eq:mfp}). We find that, as calculated with PS methods,  theoretical VCD spectra directly match experimental spectra, sometimes outperforming the standard MFP theory considerably.  Altogether, the data in Fig. \ref{figvcd}  offer another strong vindication of phase-space electronic structure methods. For more details, see Refs. \citenum{duston:2024:jctc_vcd,zhentao:2024:jcp:vcd_basis_free}.

\section{Molecules and Materials With Spin-Orbit Coupling}
\label{sec-spin}
In Sec. \ref{sec-ps-success} above, we found  that PS electronic structure theory offers consistently improved results over BO theory, sometimes with major relative corrections. That being said, in  total absolute magnitude, for most molecules (including  oxirane  in Fig. \ref{figvcd}), the total PS correction is small -- which is not surprising given the large mass difference between electrons and nuclei. Typically speaking, for the oxirane molecule and most other molecules, the phase-space potential energy surfaces usually resemble the BO potential energy surfaces in the $\bX$ direction plus a quadratic valley (corresponding to $\bP^2/2M$) in the $\bP$ direction.

At this point, however, it must be noted that we have not yet discussed spin (or more broadly systems with degeneracy). The most exciting application of phase-space electronic structure theory concerns
molecular or material system with degenerate spin degrees of freedom and spin-orbit coupling.   From a bird's eye view, one might expect to find such an application because, in atomic units, the spin-orbit coupling has an order of magnitude determined by the square of the fine structure constant times a nuclear charge (say, $1/137^2 \times 6 = 4\times10^{-4}$ for carbon) and the change in energy for the ground state is estimated to be as $\sim \alpha^2 Z_{eff}^4 / n^3 \approx 4 \times 10^{-4}$ accounting for a carbon valence electron; vice versa, the error in BO theory is of the magnitude of $\bP \cdot \bd/M_n$, and noting  that (from Eq. \ref{eq:constrain1}) $-i\hbar \bd \approx \bp$, it follows that the BO error is roughly $ m_e/M_n $ in atomic units (which is roughly $6 \times 10^{-5}$ for carbon). 
The fact that these two small numbers are close together (with the BO parameter a bit smaller) would indicate  perturbation theory within a  BO framework will not be accurate for spin-problems if one allows the nuclei to move.
Indeed, in such a case, the phase-space potential energy  surfaces do {\em not} resemble the corresponding BO potenial energy surfaces.  

\begin{figure*}
    \centering \includegraphics[width=\linewidth]{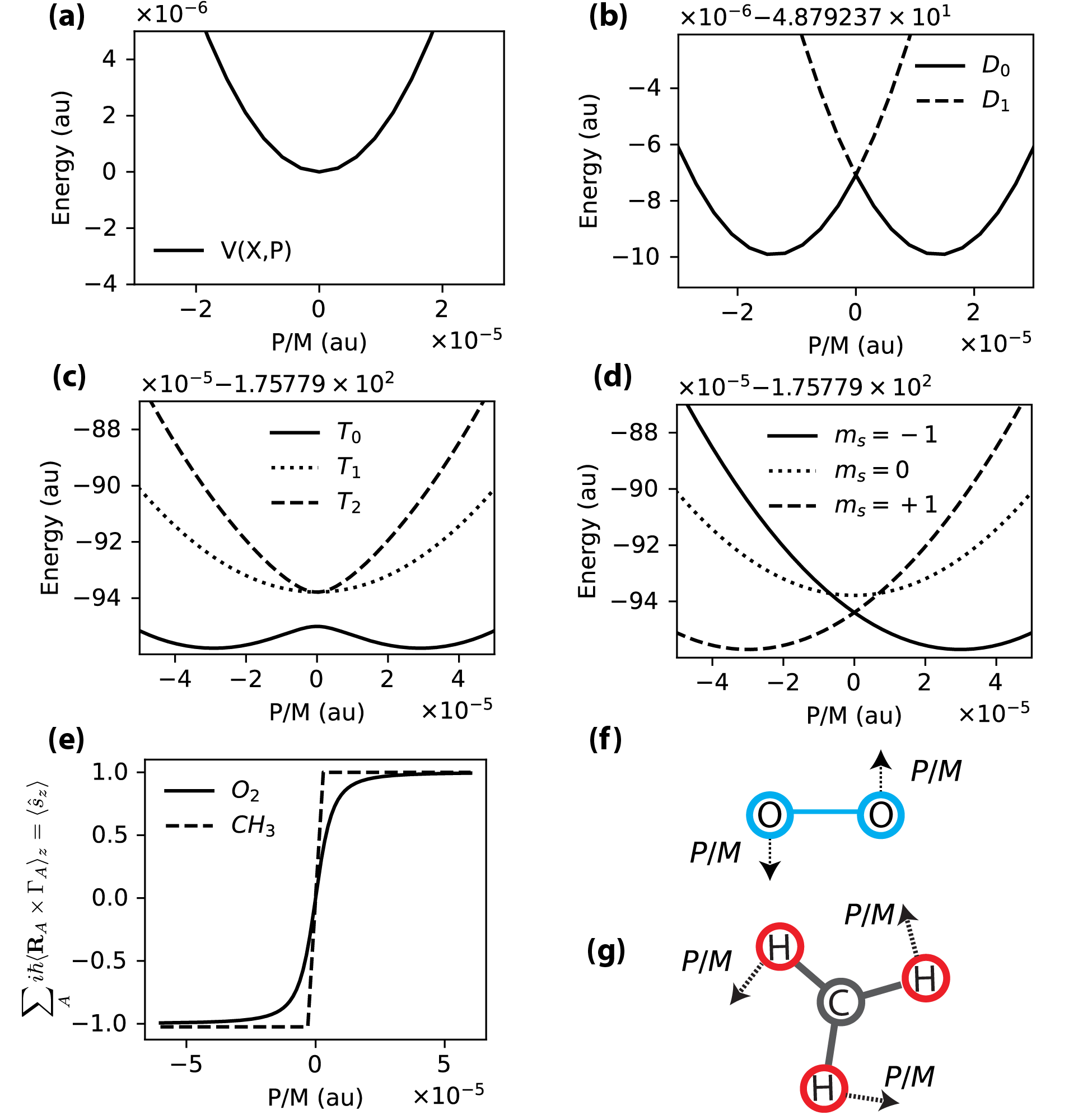}
    \caption{
    (a) The typical dependence of $V(\bX,\bP)$, namely $V(\bX) = V_{
    \rm el}(\bX) + \bP^2/2M$. (b)     The {\em ab initio} phase-space potential energy surface for the CH$_3^{\bullet}$ doublet; 
    note that the degenerate ground state has now split into two separate parabolas, i.e. broken symmetry surfaces appear.  
   (c) The {\em ab initio} phase-space potential energy surface for the O$_2$ triplet; 
    note that the degenerate ground state has now split into two three separate broken symmetry parabolas. 
    (d) The relevant spin diabats for the triplet states. 
    (e) The electronic angular momentum and electronic spin as a function of $\bP$,  demonstrating that, because of nuclear-spin angular momentum coupling, the nuclear coordinate $\bP$ is acting as a surrogate order parameter for both forms of angular momentum (orbital and spin).  
    (f) A graph of the O$_2$ and and CH$_3^\bullet$ molecules (in the $xy$-plane), delineating 
    the direction of $\bP$ that breaks the symmetry of the two spin states. Data from  Ref.\citenum{bradbury2025symmetry}.
    }
  \label{fig_p_order_s} 
\end{figure*}

To demonstrate why spin effects are paramount for phase-space electronic structure methods\cite{bradbury2025symmetry}, see Fig. \ref{fig_p_order_s}. In panel (a), we plot a typical form for a phase-space adiabat as a function of $\bP$ -- the minimum is at $\bP = 0$.  In panel (b), however, we plot the the functional form of the two lowest surfaces for a radical doublet system (CH$_3^{\bullet}$); here we find a double well appears in the ground-state.  This double minimum has a direct analogy to the band splitting that is well-known in band structure theory within the solid state world without time reversal or inversion symmetry \cite{bychkov1984properties, vanderbiltbook}.  For such a system, with an odd number of electrons and Kramers degeneracy, the ground and first excited states must be degenerate at  $\bP =0$, where the electronic phase-space Hamiltonian is time reversible, but otherwise, the spin states can be nondegenerate. 
In panel (c), we plot 
the surfaces for  triplet O$_2$.  For this case, with an even number of electrons and a ground-state triplet, there is zero field splitting and often no degeneracy at all. This form for the phase-space adiabatic surfaces can be easily rationalized in terms of the spin diabats drawn in panel (d). 

Let us now examine exactly how a double well emerges.  If we minimize $E_{\rm PS}(\bX,\bP)$ and seek a solution to $ {\partial E_{\rm PS}}/{\partial \bP} = 0$, the result is clearly:
\begin{eqnarray}
    \bP_A^{eq} = \langle \hbGamma_A \rangle_{eq}
\end{eqnarray}
where $\langle\hO\rangle_{eq} = \bra{\Psi_{\rm PS}(\bm X^{eq}, \bm P^{eq})}  \hO \ket{\Psi_{\rm PS}(\bm R^{eq}, \bm P^{eq})}$. 
Moreover, given Eqs. \ref{eq:constrain1} and \ref{eq:constrain3} above, it then follows that:
\begin{eqnarray}
\label{eq:eqp}
    \sum_A \bP_A^{eq} + \langle \hbp \rangle_{eq} = 0 \\ 
 \label{eq:eql}
    \sum_A \bX_A^{eq} \times \bP_A^{eq} + \langle \hbl + \hbs \rangle_{eq} = 0 
\end{eqnarray}
We conclude that, whenever there is a nonzero electronic spin at equilibrium, then unless $\langle \hbl + \hbs \rangle_{eq} = 0$, we must have $\bP_A^{eq} \ne 0$. Moreover, if we find one solution  to Eq. \ref{eq:diagHPS} with wavefunction $\ket{\Phi}$ at $(\bX,\bP)$, we must also expect another solution by applying the time reversal operator  at $(\bX,-\bP)$ with wavefunction $\mathcal{T} \ket{\Phi}$ (where $\mathcal{T}$ is the anti-unitary time reversal operator).  
Here, $\bP_A^{eq} \ne 0$ is the canonical momentum (which should not be confused with the kinetic momentum $\bPi_A^{eq}$).  At the equilibrium point, the kinetic momentum is indeed zero,  $\bPi_A^{eq} = 0$; the disagreement between these momenta merely enforces the fact that the {\em electrons} have nonzero momentum at equilibrium.
Effectively,  the nuclear momentum becomes a well-defined reporter of electronic spin, so that  the
potential energy surface for a doublet or triplet  looks like Fig. \ref{fig_p_order_s}(b-c) as a function of $\bP$. In this context, we are strongly reminded that spin is a form of angular momentum (and the total angular momentum is conserved), which is not always taken into account in modern electronic structure theory.   

\begin{figure*}[t] 
    \centering
    \includegraphics[width=\linewidth]{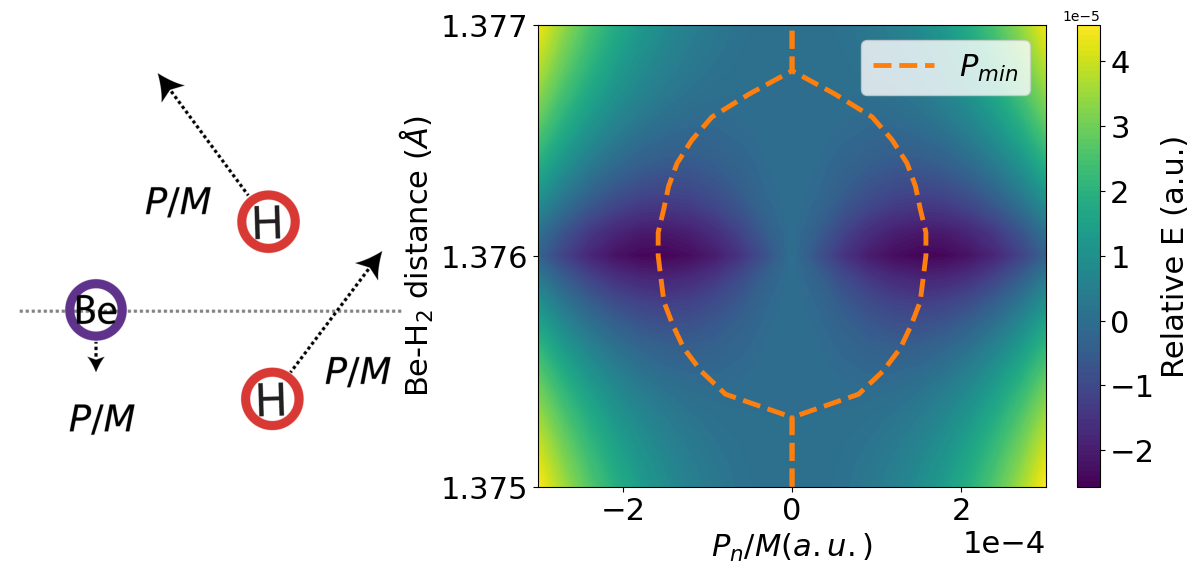}
    \caption{ A contour plot of the phase-space surface $E_{\rm PS}(\bX,\bP)$ of BeH$_2$ (relative to $E_{\rm PS}(\bX,0)$)  in the vicinity of a conical intersection. For the spatial coordinate ($y$-axis), we use Eq.~\ref{eq:beh2_path} above. For the momentum coordinate ($x$-axis), we plot the velocity vectors shown on the left. Data from Ref. \citenum{duston2025ci}.  Note that for this problem, a double minimum occurs {\em without} any spin-orbit coupling but rather as a result of mixing spatially different states through nuclear-electronic orbital coupling.}
    \label{fig:tito:beh2}
\end{figure*}

At this point, there are then two fundamental questions remaining. The first question is: Do double minima arise only for spin systems? The answer is no. In Fig. \ref{fig:tito:beh2}, we plot the potential energy surface for the BeH$_2$ molecule along the well-known following path\cite{headgordon:2015:CRHFstability}
\begin{eqnarray}
\label{eq:beh2_path}
\text{Be}&:&(0,0,0)\nonumber\\
\text{H}&:&(x,1.344-0.46x,0)\\
\text{H}&:&(x,-1.344-0.46x,0)\nonumber
\end{eqnarray}
that crosses through a conical intersection (CI). Note that, near the conical intersection, there is  a double minimum in $\bP$.  In other words,  a single path leading directly through the conical intersection is deformed into two paths on either side of the CI, each path carrying a different amount of electronic angular momentum. (Interestingly, note 
that a CI in phase space has a branching plane of dimension three rather than two\cite{martinez:2006:ci_topology_wrong}).

The second question that must then be posed is: 
What is the size of the barrier between broken symmetry states? 
For the CH$_3$ molecule, according to Fig. \ref{fig_p_order_s}, we find a barrier of magnitude $3\times 10^{-6}$ au which is quite small and difficult to measure.  One can rightly wonder whether the presence of such a small barrier has any physical meaning. That being said, despite the small size of the barrier for CH$_3$ radical, a little more thought reveals that the double minima can be much larger. After all, for a magnetic system, flipping a spin can require a spin-exchange energy on the order of 0.1 electron volts\cite{stanton:2020:exchange_couplings}. Moreover, as we will discuss in Sec.  \ref{sec-edh}, with the proper statistical mechanics interpretation, the present approach should also be capable of modeling a macroscopic Einstein-de Haas effect\cite{einstein_dehaas_1915a,einstein_dehaas_1915b} (with a macroscopically large energy barrier between wells).

\section{A Plethora of Macrosocopic Applications With Spin} \label{sec-app}

Above, we have outlined the power of phase-space  theory to give an extremely new perspective on how spin couples to electronic and nuclear motion within electronic structure calculations. Let us now consider different future applications.

\subsection{The Einstein-de Haas Effect}\label{sec-edh}
Modeling the Einstein-de Haas effect is a clear target application for phase-space electronic structure theory.  As a reminder, the Einstein-de Haas effect results when an oscillating magnetic field is applied to a chunk of a ferromagnet. In the most simple case, imagine that the applied field is set to a constant $\bB$ for time $T$, then $-\bB$ for another time $T$, then $\bB$ for time $T$, etc. 
Now, when a material is placed in the presence of a constant magnetic field (say in the $z$ direction), it is straightforward to prove that angular momentum in the $z$ direction will be conserved\cite{bhati:2025:jpca:magnetic_part1,greenshields:2014:prl:angular_momentum,johnson:1983:rmp:magfields,cederbaum1991review}.
The Einstein-de Haas effect is the phenomenon whereby, as the spins align but the total system must conserve angular momentum, the entire chunk of a ferromagnet begins to rotate  (say, clockwise for  time $T$, then counter-clockwise for another time $T$, 
then counter-clockwise for another time $T$, etc).

Einstein-de Haas cannot be easily described within BO theory and has motivated a host of new approaches\cite{lemesheko:2015:prl,lemesheko:2016:prx,lemesheko:2017:prl}. 
That being said, the phenomenon above should be easily describable within a phase-space electronic Hamiltonian.  To understand why, note that the stable states of a ferromagnetic system with $N$ atoms support  macroscopic spin polarization, $\langle\hbs\rangle = N {\bm s_0}$, where ${\bm s_0}$ is the spin polarization per atom.  Presumably, there may or may not also be some orbital polarization per atom, ${\bm l_0}$.  If we define $\lambda_A$ to satisfy
\begin{eqnarray}
    \bX_A^{eq} \times \blambda_A^{eq} + \hbl_0 + \hbs_0 = 0 
\end{eqnarray}
In such a case, at the equilibrium geometries characterized by Eq. \ref{eq:eql}, 
one expects a macroscopic nuclear displacement $\bP_+ =  (\bP_1, \bP_2, \ldots, \bP_N) = (\blambda_1, \blambda_2, \ldots, \blambda_N)$. At the same time, of course, we would expect a macroscopic polarization of $\langle\hbs\rangle = -N {\bm s_0}$ if we chose $\bP_- = (-\blambda_1, -\blambda_2, \ldots, -\blambda_N)$.  More generally,   we must expect to find roughly $2^N$ minima  at momenta  $(\pm \blambda_1, \pm \blambda_2, \ldots, \pm \blambda_N )$ with expected spin $\langle\hbs\rangle$ satisfying  $ - N {\bm s_0} < \langle\hbs\rangle < N {\bm s_0}.$ 

\begin{figure*}
    \centering
\includegraphics[width=0.9\linewidth]{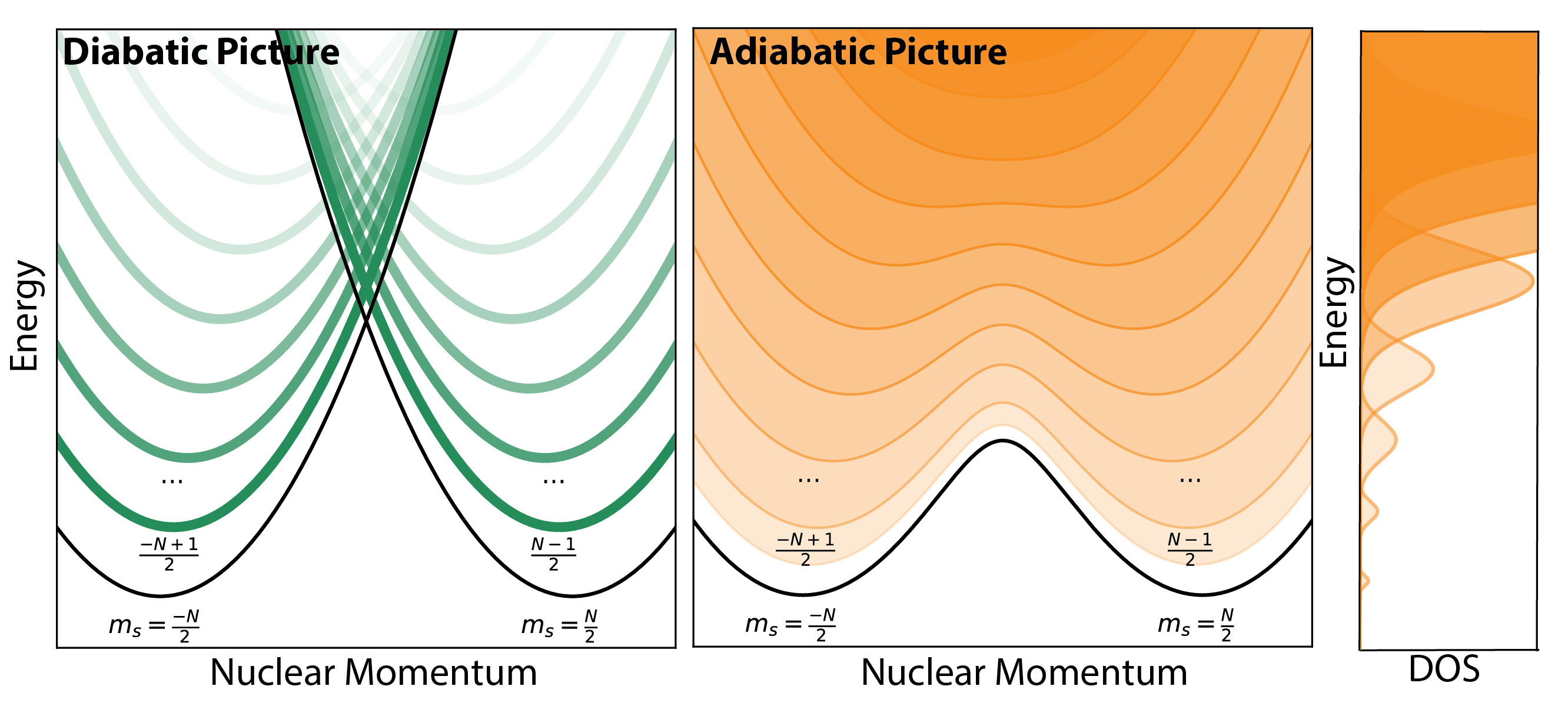}
    \caption{ A hypothetical model of the potential energy curve in $\bP$-space, along which dynamics move during the Einstein de Haas experiments. Although there are likely $2^N$ minima, here we sketch the energy along the line connecting $\bm P_+$ and $\bm P_-$ , the two positions of the global minima.  Note that, because a PS approach correctly couples  angular momenta together, the nuclear momentum $\bm P$ acts as a surrogate order parameter for spin. On the left, we plot the diabats (labeling the spins) and in the middle we plot the adiabats.  Note also that the barrier between basins should be macroscopically large here.   
    In the right hand panel, we plot  the proposed density of states (DOS), which must grow larger with larger energy. For more details, see Ref.\citenum{bradbury2025symmetry}}
    \label{fig:edh}
\end{figure*}

Given the enormous number of possible spin states, one instructive next step is to walk along the line between  $\bP_-$ and $\bP_+$. Along such a one-dimensional path in momentum space (that serves as a reaction coordinate), one can plot the ground state potential energy surface. For this problem, one will undoubtedly find a macroscopically large energy barrier in $\bP-$space between the two macroscopically polarized spin states ($-N{\bm s_0}$ and $N {\bm s_0}$) with a macroscopically large barrier between them. See Fig. \ref{fig:edh}(a). Along such a path, one can also calculate the  expectation value of the spin $\langle\hat {\bm s}\rangle$ (as hypothetically sketched in Fig. \ref{fig:edh}(b)). 
Now, of course, dynamics certainly do not follow such a one-dimensional path; $N$ spins never invert at the same time.  Nevertheless, such a one-dimensional graph should be enough to inform us as to how much angular momentum is transferred  to the nuclear degrees of freedom (instead of orbital electronic angular momentum) after a macroscopic spin flip.  In other words, this approach should be enough to tell us about the final angular speed of the metal substrate after switching the macroscopic spin. After all, with a phase-space electronic Hamiltonian, one is guaranteed to conserve the total momentum.

Finally, we have sofar discussed only adiabatic motion along the ground state and we have not mentioned anything about entropic effects or more generally dynamics in the presence of so many curve crossings (e.g., what is the time scale  required to start rotating in the opposite direction after we switch the magnetization?).  To make progress, in principle, one should really keep track or all $2^N$ states. 
To that end, if one wishes to include entropic effects, then  just as for the Ising model, one can also generate a free energy surface as a function of this one-dimensional coordinate by sampling over different nuclear configurations; one could also measure the average  $\langle\hat{\bm s}\rangle$ by sampling over many configurations.

Next, consider dynamics, which evolve on potential energy surfaces (not free energy surfaces). In principle, one would like to  simulate nonadiabatic dynamics in the full $6N$-dimensional nuclear phase-space.
That being said, in a coarse grained sense, one can  imagine trying to learn something about nonadiabatic dynamics along the reduced one-dimensional reaction coordinate above.
Here, we imagine the relevant curves must look appear as they do in Fig. \ref{fig:edh}(c), where the red arrow captures the amount of spin-exchange energy that can be harnessed by aligning spin, and the blue arrow captures the amount of energy that is tied to nuclear-spin angular momentum coupling (including indirectly the amount of energy stored in electronic orbital angular momentum).  With such a figure, one can begin to contemplate possible approaches to the present nonadiabatic problem. For instance, one might consider attempting to use models of ``electronic friction''\cite{tully:1995:electronic_friction,suhl:1975:prb_elfriction,dou:2017:prl} to treat the resulting nuclear dynamics in the presence of so many electronic states. 
At the end of day,  
simulating the Einstein-de Haas effect with a phase-space electronic Hamiltonian represents and immediate target for future dynamical studies.



\subsection{Chiral Induced Spin Selectivity}
While the Einstein-de Haas effect was predicted and observed one hundred years ago\cite{richardson:einstein_dehaas,einstein_dehaas_1915a,einstein_dehaas_1915b}, 
there is today a very modern demonstration of dynamics involving angular momentum that remains
unexplained, namely the  observation of chiral induced spin selectivity (CISS)\cite{bloom:cissreview:2024:chemrev}. In its most simple form, the CISS effect was originally observed in 1999 by Waldeck and Naaman\cite{naaman:waldeck:1999:original_ciss} as far as the observation of asymmetric scattering of polarized electrons in thin films. In 2011, a seminal experment in photoemission by Zacharias {\em et al}\cite{naaman:2011:science:ciss_dna} demonstrated conclusively that there was a spin preference in the emission of electrons through a chiral layer of DNA.  Since 2011, there continues to be a frenzy of activity in this field, with an enormous number of reports of CISS. Apart from photo-emission experiments in the vein of Ref. \citenum{naaman:2011:science:ciss_dna},  the other most common measurements are spin-valve conduction experiments\cite{adhikari:pengxiong:2023:ciss_spin_valve_topology} where strong magnetic field effects are observed.  

There are other experimental manifestations as well\cite{bloom:cissreview:2024:chemrev}; the list of references above is far from exhaustive and reflects the ignorance of a theoretical group that is slightly removed from the actual experiments. 
Interestingly, though, it is clear there are quite a few contradictions in the literature.\cite{dubi:2023:temperature_review_ciss} For example, some studies report that CISS signal decreases in magnitude with increasing temperature\cite{yang:ciss:temp_inc_ciss_dec:2023:nat_chem,naaman:2013:pccp:temp_inc_ciss_dec}. Other studies, predict that the CISS effect will increase with increasing temperature\cite{naaman:2022:ciss_temperature_increase_ciss_increase}. 

At present, there is no clear explanation for CISS.\cite{evers:2022:ciss_theory_review} Early attempts to rationalize CISS hypothesized that a conducting electron 
moving through a helix must interact with its own magnetic field, but these effects are now known to be small\cite{gersten_nitzan:2013:jcp_chiral}. Broadly speaking, this failure leaves three standing hypotheses. First, 
Nitzan and Gersten\cite{gersten_nitzan:2013:jcp_chiral,binghai:2021:nature_materials:ciss} 
have suggested that electronic transmission through a chiral medium selects for a given electronic orbital angular momentum,
and that electronic orbital angular momentum is tied to electronic spin through spin-orbit coupling.  Second, there is also a related hypothesis that a magnetic moment appears at the interface 
(the so-called ``spinterface'')\cite{dubi:2021:jacs:spinterface} 
and selects for a given spin angular momentum.  Here, we distinguish between these two hypotheses insofar as orbital-spin locking should be possible not only at an interface; vice-versa,  orbital-spin locking is only one possible means to create a spinterface. Third and finally, it has been argued by some (including the authors) that CISS might arise from the interaction of conducting electrons colliding with phonons given the need to conserve angular momentum\cite{fransson:2020prb:vibrational,fransson:2022:jpcc:vibrations,wu:2020:neartotal,bian:2021:perspective}.

In any event, none of these mechanisms would appear satisfactory to explain all of the CISS behavior in the literature.   First, one expects orbital-spin locking to depend on the size of the spin-orbit coupling; however, there are photo-emission experiments with heavy and light metal substrates that reveal no such dependence.\cite{zacharias:2018:jpcl:helicene_monolayers,fontanesi:naaman:2013:pna:ciss_aluminum_still_effect}
 Second, the spinterface mechanism is not compatible with the observation that, when running conduction experiments through molecular chairs,  the spin effect is enhanced when longer and longer the polymer\cite{naaman:2011:nanoletters:dna,dubi:2022:spinterface:length_dependence}. 
Third, it would appear unlikley that nuclear vibrations can play a role in photoemission experiments\cite{naaman:2011:science:ciss_dna}, given that electrons are ejected almost instantaneously.
For the reasons cited above, there is  today
a suspicion among many theorists that there may be several underlying mechanisms for CISS.
To that end, we will now posit that the three explanations above can in fact be rationalized within a phase-space framework, 
and that future phase-space calculations should indeed be able make predictions therein.

\subsubsection{Nonequilibrium (non-boltzmann) Initial Conditions}
We have noted above that for cases with spin degrees of freedom, one should expect broken symmetry minima in the momentum ($\bP$) coordinate.
With this in mind, there is a clear possibility that, for systems exhibiting CISS on a metal surface, the dynamics being measured are initialized with a non-equililbrium distribution of spin at the onset. After all, there is no reason to presume that the deposition and/or adsorption of chiral molecules on metals surfaces need proceed in a time-reversible fashion and indeed, there is some evidence  that such deposition is dominated by kinetics and has no thermodynamic basis. \cite{buergler:2024:ciss_adsorption_kinetic_controlled,waldeck_subotnik:2023:jpcc}  Now, if the adsorption process 
were to lead to molecules attaching to a surface with a preference for one spin state, this scenario would perhaps correspond to the spinterface mechanism above.

One means of checking this hypothesis would be to apply and remove different magnetic fields to a non-magnetic system and repeat the experiment over and over to reach equilibrium.  For instance, in the photoemission experiments on gold of Ref. \citenum{naaman:2011:science:ciss_dna}, one could apply  a magnetic field in the $+z$ direction for several minutes, remove the magnet, and then measure again; thereafter, one could repeat the exact same experiment in the $-z$ direction.   Alternatively, one could heat and cool the system repeatedly. The ansatz  would be that, if one  stirred up the system enough to overcome barriers, one might find that the photoemitted electrons lose their spin-polarization as the system now starts from a truly time-reversible equilibrium Boltzmann ensemble. 
One would also expect that, upon increasing the temperature, there would either be no effect or a {\em decrease} in the spin signal (depending on the change of temperature vs. the barrier height between different $\bP$-minima).

\subsubsection{Electronic Orbital-Spin Locking and Photo-Emission}
In Fig. \ref{fig_p_order_s} above, we showed a potential energy surface in phase space with broken symmetry 
minima arising from  a spin-coriolis force and the presence of degenerate spin states; 
at each stationary point in phase space, the electronic state 
has a well-defined spin direction. Vice versa, in Fig. \ref{fig:tito:beh2} above, we showed a broken symmetry potential energy surface 
that arises from an orbital coriolis coupling (without any regard for spin) and the presence of two degenerate electronic
states at a conical intersection; for this system, at a stationary point in phase space, the electronic state 
has a well-defined electronic orbital angular momentum. 
Now, for an open shell system residing on a
metal surface, one would expect to find  degenerate states because of both spin degeneracy and  electronic degeneracy (on the metal). Therefore, one would expect to find broken symmetry solutions with both well-defined spin and orbital angular momentum directions at the same time, i.e.   orbital-spin locking that is consistent with  Eq. \ref{eq:constrain3} above. For photo-emission experiments that do not depend on state preparation, this mechanism\cite{gersten_nitzan:2013:jcp_chiral} would appear to be relevant. Note that the extent of orbital-spin locking will depend on several factors, including the orbital and spin coriolis forces, and  need not depend only on spin-orbit coupling.  In the future, {\em ab initio} calculations of orbital-spin locking should allow for a test of this proposed mechanism.

\subsubsection{Coupled Electronic-Nuclear Dynamics and Marcus Theory}
Lastly, consider again the usual diabatic picture of the donor and acceptor states in Fig. \ref{fig_problems}(c).  The key insight of Marcus theory was that electron transfer is gated by nuclear motion; moreover, because nuclei are heavy and electrons are light,  a change of electronic state cannot occur at the same time as a change in the nuclear state. Thus, in order to conserve {\em energy}, an electron can transfer only at the crossing point in Fig. \ref{fig_problems}(a), where the two diabatic electronic states have the same energy. Note that, within this Marcus picture, there is no notion of the need to conserve {\em momentum}.

\begin{figure}[h]
    \centering \includegraphics[width=\linewidth]{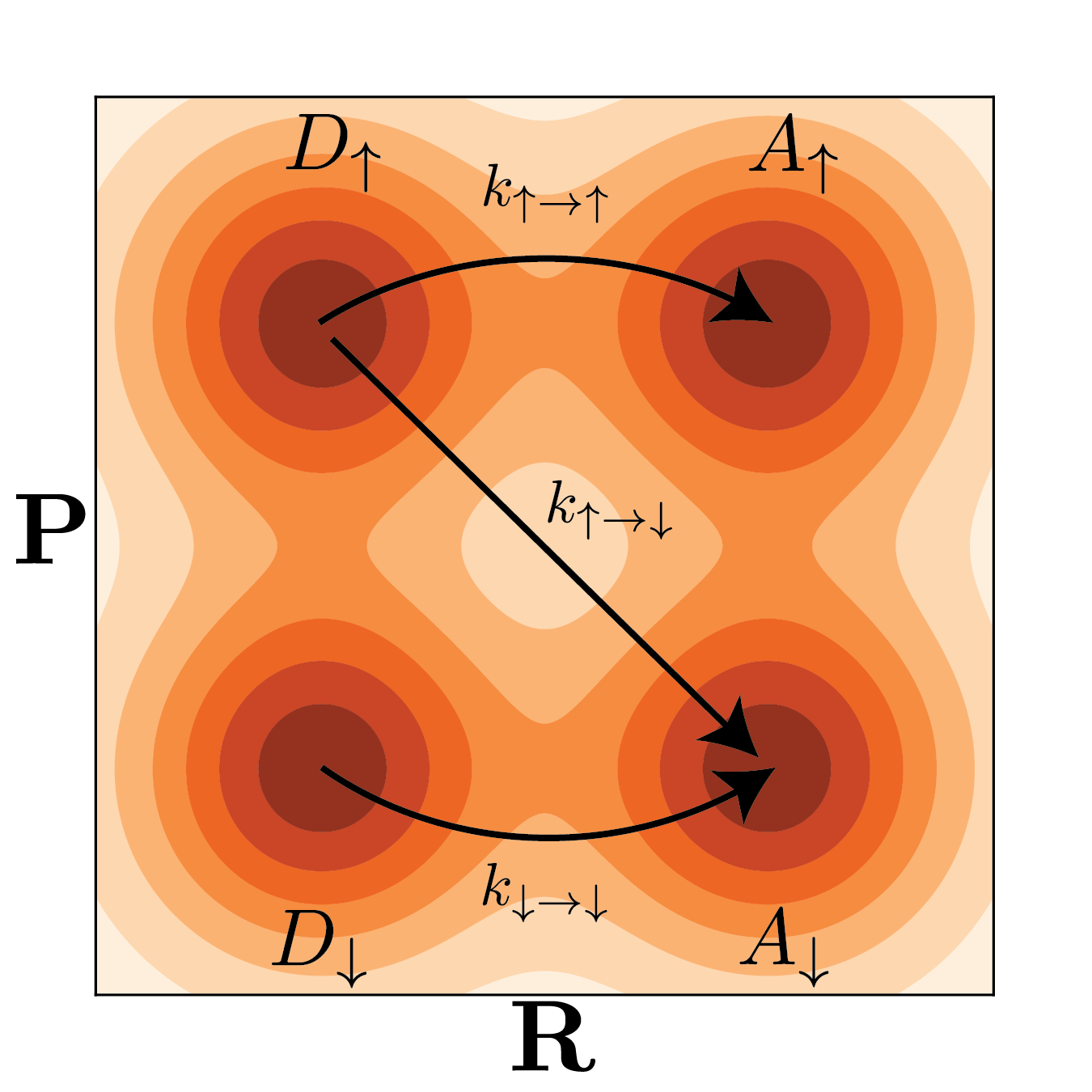}
    \caption{Within a spin-dependent Marcus theory that conserves momentum and energy, one must imagine that there are four wells at ($\pm R_0, \pm P_0)$. As indicated on the graph, there are three different classes of rates that can be identified and measured:  spin conserving electron transfer (ET) rates ($k_{D\uparrow \rightarrow A \uparrow}$ and $k_{D\downarrow \rightarrow A \downarrow}$), spin non-conserving ET rates ($k_{D\uparrow \rightarrow A \downarrow}$ and $k_{D\downarrow \rightarrow A \uparrow}$) and spin-flip rates (e.g., $k_{D\uparrow \rightarrow D \downarrow}$). In general, if the rate prefactors for  $k_{D\uparrow \rightarrow A \uparrow}$ versus $k_{D\downarrow \rightarrow A \downarrow}$  are very different, one would expect the total rate difference to increase with temperature; moreover, a preference for one spin should be amplified for systems with more than two sites.}
    \label{fig3or4parab}
\end{figure}

The notion of momentum conservation never arises in standard Marcus theory; effectively,  Marcus theory ignores electronic momentum insofar as the total  momentum is the nuclear momentum.   While this failure of Marcus theory is usually hidden when discussed in the classroom,  a perceptive student can understand the problem in two ways. $(i)$ First, if one imagines constructing the Marcus diabatic states from BO states as in Fig. \ref{fig_problems}, then far from a crossing the Marcus diabats agree with BO states; and, as we have already discussed in Sec. \ref{sec-BO-fail2}, BO theory ignores electronic momentum. $(i)$ Second, note that Marcus theory assumes the existence of two exactly diabatic states.  Historically, this feature was understood to enforce the notion that electronic dynamics are very fast (instantaneous) compared with nuclear dynamics.  However, it is well known that exactly diabatic states do not exist for multidimensional problems unless the Berry curvature in Eq. \ref{eq:nonad:berry_curvature} vanishes\cite{mbaer:1975:cpl,mead:1992:rmp,littlejohn:2022:jcp:parallel}.  More generally, if one wishes to recover electronic momentum, one must account for dynamic correlation and correlate a given electronic state with many other electronic states (as in Eq. \ref{eq:expand} above). No matter how one approaches the problem, the conclusion is clear:  Marcus theory ignores electronic momentum and/or momentum conservation.

 If one wishes to include electronic momentum within Marcus theory, and run dynamics that conserve the total momentum, the natural approach would be to use a phase-space electronic Hamiltonian (and construct diabats from the resulting electronic surfaces). For an electron transfer between donor (D) and acceptor (A) with spin (i.e.  doubly degenerate diabats), given the discussion in Sec. \ref{sec-spin}, we would now expect to find four electronic states of interest (i.e. minima in phase-space):
 $\ket{D \uparrow},\ket{D \downarrow},\ket{A \uparrow},\ket{A \downarrow}$.
corresponding to four distinct minima in phase-space (as shown in Fig. \ref{fig3or4parab}.)
The job of the theorist is then to  evaluate the kinetics between these four wells, some of which correspond to electron transfer, some spin flipping, some both.  In the future, one can imagine running molecular dynamics simulations (perhaps with phase-space surface hopping\cite{wu:2022:pssh,yanzewu:2024:jcp:pssh_conserve,bian:2022:pssh,bian:2024:pssh_translations_rotations} or Ehrenfest\cite{coraline:2024:jcp:Ehrenfest_conserve} dynamics) to account for nonadiabatic transitions; one can also imagine running fermi golden rule (FGR) calculations,  noting that an equilibrium two-state FGR will necessarily need to include more than two states or nonequilibrium effects for a meaningful answer\cite{chandran:2022:jpca:nefgr}.  Of course, the size of the SOC will be paramount in these calculations, but equally important will be the shape of the surfaces especially near the crossings seam. For maximal impact, it will be necessary to generate some of these surfaces with {\em ab initio} methods; it may also be interesting to consider how these surfaces and crossings can be altered explicitly by an applied magnetic field\cite{zimmt:1983:magnetic_review}.

The discussion above regarding the spin dependence of electron transfer should have direct consequences for interpreting CISS experiments. To that end, consider conduction through a large system with many molecular fragments (e.g.  polymers and/or long molecular chains).  For such systems, let us imagine there are $N$ monomers (with $2^N$ minima in phase-space [just as for the Einstein-de Haas effect]). Conduction involves a system traversing all of these minima in sequential order and for such a process, and if there is a spin preference in site-to-site electron transfer, the present mechanism would likely explain
why the  CISS effect seems to grow linearly for longer molecular chains\cite{naaman:2011:nanoletters:dna,dubi:2022:spinterface:length_dependence}. As far as this vibrational mechanism is concerned,
one would predict that the observed spin polarization should {\em increase} as the temperature increases  and nuclear motion becomes more and more important (and the electronic conduction is dominated by hopping rather than band transport).   Ideally, one will be able to predict a measurable isotope effect within this framework for experimental confirmation.

\subsection{Magnetic Field Effects More Generally}\label{sec-mag}
Before concluding, it is worth emphasizing that, thirty years after Steiner's influential review of magnetic field effects\cite{steiner:1989:magnetic_review}, theoretical chemistry still faces enormous difficulties with respect to modeling magnetic field effects (MFEs) -- a vast number of MFEs still cannot be explained today.  The rub of the matter is that external magnetic fields are extremely weak such that their effect on a BO potential energy surface is extremely limited, so that developing a theory of magnetic field effects based on BO surfaces appears very unnatural.   
To that end, Cederbaum \cite{cederbaum:1988:pra:magnetic_fied_pseudomomentum_born_oppenheimer,cederbaum:1989:pra:magnetic_fied_pseudomomentum,cederbaum:1997:ijqc:magnetic_fied_thoughts}) pointed out long ago that magnetic field effects are notorious for promoting non-Born Oppenheimer effects. The reasoning is quite clear: both electrons and nuclei should  experience a magnetic field, but charged particles feel magnetic fields only during motion.  Now, within the BO approximation, one first freezes the nuclei, and then solves for the stationary states of the electron in the presence of the $\bB$-field. Thereafter, the nuclei move along the BO surface and experience the same magnetic field. The problem is then that,  when the nuclei start moving, they also drag electrons with them, such that the electrons should again experience the magnetic field.  However, such an effect is not included in BO dynamics, where the electronic states were already calculated (as a function of only the nuclear position $\bX$) and do not carry momentum!

Obviously, a phase-space approach is one means to ameliorate the problem above. Namely, if one parametrizes the electronic states by both nuclear position $\bX$ and nuclear momentum $\bP$, one can include both (i) the direct effect of the magnetic field on the electronic momentum through a $\hbp \cdot \bA(\hbx)$ term (where $\hbp$ is the electronic momentum operator and $\bA$ is the external magnetic field vector potential ) as well as $(ii)$ the indirect effect of the  magnetic field on the electronic momentum as induced by nuclear motion through a $\hbp \cdot \bPi(\bX)$ (where $\bPi(\bX)$ is an effective nuclear momentum operator). Indeed, in a recent set of papers\cite{bhati:2025:jpca:magnetic_part1,bhati:2025:jpca:magnetic_part2}, we have hypothesized the form of a phase-space electronic Hamiltonian as relevant for molecules and materials. Here, many nuances arise, including especially the fact that, during the resulting dynamics, one conserves not the total momentum but rather the total pseudomomentum\cite{helgaker:2022:jcp_conservation_laws_magnetic_field} (whose value depends on the choice of magnetic origin). Nevertheless,  in Refs. \citenum{bhati:2025:jpca:magnetic_part1,bhati:2025:jpca:magnetic_part2}, we have developed one possible phase-space electronic Hamiltonian which satisfies many key targets: the Hamiltonian is gauge invariant, adiabatic dynamics  conserve the total pseudomomentum in all directions as well as the angular momentum in the direction of the magnetic field, the energy is translationally invariant, the electronic energies are exact for a hydrogen atom, and the phase-space Hamiltonian reduces to Eq. \ref{eq:HPS} in the absence of a magnetic field. One would hope that such a phase-space electronic Hamiltonian will be helpful as far as explaining magnetic field effects in the future, especially organic magnetoresistance\cite{gobbi:2017:oranic_magnetoresistance_review}.

\section{Conclusions and Outlook}
Our discussion above represents the beginning of a long journey into new aspects of electronic structure theory and chemical dynamics and almost certainly just the tip of a big iceberg. After all, almost all of our theoretical chemical intuition is based on BO potential energy surfaces, $V_{\rm BO}(\bX)$; and yet, the BO potential energy surface represents just one strip of the phase-space potential energy surface $V_{\rm PS}(\bX,\bP)$ 
i.e. $V_{\rm PS}(\bX,0) = V_{\rm BO}(\bX)$. Thus, given how strongly the BO perspective has dominated our conception of chemistry heretofore, it remains to explore what we will learn by exploring the phase-space energy surface for $\bP \ne 0$. Interestingly, as a side note,
if we adopt a phase-space point of view (as a substitute for a BO framework), one casualty within our graduate education will
the clear distinction between thermodynamics and dynamics.  Indeed, if electronic structure is performed in a moving frame, it becomes impossible to clearly differentiate between static equilibrium distributions and dynamical trajectories. Adopting one's point of view to a phase-space electronic structure Hamiltonian will require some adjustments.  

Intuitively, one must wonder why a phase-space perspective on electronic structure theory has not been explored before by quantum chemists and solid state physicists. Presumably, the biggest drawback to the present theory is that, if one wishes to explore quantum dynamics with quantum nuclei, invoking Wigner transforms (which are not regularly taught to graduate students) might appear awkward and an unnatural framework towards an exact solution.   That being said, as discussed around Eq. \ref{eq:hps:expand} above, one can formally expand the total Hamiltonian in powers of $\hbar$ within a phase-space framework and construct exact eigenstates.

Looking forward, in order to fully explore those areas of chemistry where BO fails, from a  practical perspective, many innovations will be needed (as far as implementation is concerned).  Electronic structure theory has made enormous progress over the last 50 years\cite{szabo:ostlund,helgakerbook}, but almost all based on the BO perspective.
At a minimum, future progress in phase-space electronic structure theory will demand fast electronic structure codes and rewriting large chunks of existing quantum chemistry packages. Especially important will be  the development of fast optimizers for DFT/HF in the presence of complex-valued Hermitian Hamiltonians, extensive benchmarking to best extract the atomic parameters for the $\theta(\bx)$ functions in Eq. \ref{eq:theta} (or more generally, the optimal form of the $\bf{\hat \Gamma}$ operator), and then design of optimal grids  to quickly evaluate those $\theta(\bx)$ functions. Given the urge to extend the calculations to molecular clusters
and nanosystems and explore  the broken symmetry states (with  multiple minima in momentum $\bP$), fast code (including gradients) will be essential so that we can characterize stationary points as a function of molecular size. Future work will also no doubt want to investigate phase-space electronic Hamiltonians in bulk solids. 
Finally, performing calculations in an explicit magnetic field (which is a gauge field) will require either that we use a plane-wave basis\cite{quantumespresso:2009} or that all of the implementations described above be interfaced with gauge invariant atomic orbitals (GIAOs)\cite{pulay:1990:giao:nmr,pulay:2007:giao}, which adds another layer of complexity for the future.   Indeed, there is an enormous amount of work to be done by the next generation of electronic structure theorists seeking to go beyond the BO approximation. And ideally, the steps above will also attract the interest of the next generation of quantum dynamicists, seeking  to establish spectroscopic and path integral formalisms\cite{feldman_barak_hirshberg:2023:pathintegrals_boson:jcp} using the present phase-space surfaces. We note that theoretical chemists have a long history of working with Wigner transforms to understand nonadiabatic dynamics semiclassically.\cite{martens:1997:partwig,martens:1998:partwig,kapral:1999:jcp, kapral:2009:burghardt,kelly:2010:jcp,kapral:2000:jcp, coker:2012:iterative,geva:shi:2003:semiclassical_relaxation:jpca}

Having listed some future methodological and practical considerations, let us return to the potential impact of a phase-space approach on chemistry, physics and materials. Beyond the magnetic and spin applications already listed above in Sec. \ref{sec-app}, including Einstein-de Haas and CISS, 
one must presume that there are many other physical effects (either currently known or unknown) that can be explained using phase-space (rather than BO) dynamics.  For example, there has been an explosion of interest in the CISS effect precisely because of potential to use spin statistics in catalysis (as seen, most clearly, in the water splitting example where the presence of chiral catalysts limited the production of H$_2$O$_2$ presumably by spin aligning the OH$^\bullet$ radicals\cite{naaman:2015:jpcl:water_splitting}).  Thus, one must wonder if, one day in the future, to better fit with phase-space dynamics,  organic chemist wills move dots with arrows so as to keep track of spins during new synthetic protocols.  Even better, will phase-space approaches be helpful in rationalizing new  catalytic approaches, perhaps photo-redox approaches, in magnetic fields? 
Will phase-space potential energys surfaces help us explore the plethora of magnetic field effects described in Ref. \citenum{steiner:1989:magnetic_review}? Or avian navigation\cite{hore_manolopoulos:2016:pnas:mayberadicalpair}?
As one final example, note that the $\hbGamma$ operator in Eq. \ref{eq:HPS} captures how an electron and a nucleus become entangled when near one another; if we pull out the center of mass, we find that the nucleus effectively drags the electron.  Consider now a second electron that is attracted to the nucleus; this attraction will then induce a slight attraction between the second and first electrons. Thus, one must wonder whether the present phase-space electronic Hamiltonian may teach us something new about superconductivity as well.\cite{tinkham_superconductivity}

These are exciting times.

\section{Acknowledgments}
This work was supported by the U.S. Air Force Office of Scientific Research (USAFOSR)
under Grant No. FA9550-23-1-0368 and the National Science Foundation
under Grant No. CHE-2422858.  We acknowledge the DoD High Performance Computing Modernization Program for computer time.

\bibliography{finalbib2}
\end{document}